\documentclass[%
 reprint,
 amsmath,amssymb,
 aps,
]{revtex4-1}

\usepackage{graphicx}
\usepackage{dcolumn}
\usepackage{bm}
\usepackage{comment}
\usepackage{hyperref}
\usepackage{amsmath}
\usepackage{soul}
\usepackage{listings}
\usepackage{multirow}
\usepackage{algorithm}
\usepackage{algpseudocode}
\usepackage{caption}
\usepackage{placeins}
\usepackage{mathtools}

\hypersetup{ 
    pdfnewwindow=true,      
    colorlinks=true,       
    linkcolor=blue,         
    citecolor=blue,        
    filecolor=blue,      
    urlcolor=blue        
}  

\algblock{Input}{EndInput}
\algnotext{EndInput}
\algblock{Output}{EndOutput}
\algnotext{EndOutput}

\def\be{\begin{eqnarray} &&} 
 
\def\ee{\end{eqnarray}} 

\begin{document}

\preprint{APS/123-QED}

\title{DeepRICH: Learning Deeply Cherenkov Detectors}

\author{Cristiano Fanelli}
 \email{cfanelli@mit.edu}
 
 \affiliation{Laboratory for Nuclear Science, Massachusetts Institute of Technology, Cambridge, MA 02139, USA}
 \affiliation{Jefferson Lab, EIC Center, Newport News, VA 23606, USA}


\author{Jary Pomponi}%
\affiliation{
Department of Information Engineering, Electronics and Telecommunications (DIET), Sapienza University of Rome, Italy
}

\date{\today}

\begin{abstract}

\begin{small}
Imaging Cherenkov detectors are largely used for particle identification (PID) in nuclear and particle physics experiments, where developing fast reconstruction algorithms is becoming of paramount importance to allow for near real time calibration and data quality control, as well as to speed up offline analysis of large amount of data. 

In this paper we present DeepRICH, a novel deep learning algorithm for fast reconstruction which can be applied to different imaging Cherenkov detectors. 
\\The core of our architecture is a generative model which leverages on a custom Variational Auto-encoder (VAE) combined to Maximum Mean Discrepancy (MMD), with a Convolutional Neural Network (CNN) extracting features from the space of the latent variables for classification. 

A thorough comparison with the simulation/reconstruction package FastDIRC is discussed in the text. 
DeepRICH has the advantage to bypass low-level details needed to build a likelihood, allowing for a sensitive improvement in computation time at potentially the same reconstruction performance of other established reconstruction algorithms.

In the conclusions, we address the implications and potentialities of this work, discussing possible future extensions and generalization.

\end{small}

\end{abstract}

\maketitle


\section{Introduction}

Imaging Cherenkov detectors \cite{seguinot1977photo} measure the velocity of charged particles and combined to independent measurements of their momentum are largely used for PID in modern particle physics experiments.
The pattern recognition of the rings is typically likelihood-based and requires computationally expensive simulations, hence different strategies (among which pre-computed look-up tables) have been developed to find a trade-off between time and reconstruction performance. 
A particular class of Cherenkov detectors is based on the detection of internally reflected Cherenkov (DIRC) light (see, \textit{e.g.}, \cite{stevens2016gluex}): light is contained by total internal reflection inside a solid radiator preserving its angular information until it reaches spatially segmented photon sensors, where typically rather complex hit patterns are observed. 

Machine learning (ML) algorithms are already the state-of-the-art in event and particle identification in high energy physics \cite{albertsson2018machine} but solutions based on ML for this kind of detectors just started being explored  \cite{derkach2019cherenkov}. 
\\The first DIRC detector was developed by the BaBar experiment at SLAC \cite{adam2005dirc}, and inspired other experiments (see, \textit{e.g.}, \cite{va2013progress,inami2014top,forty2009torch}) to utilize similar detectors, also in view of future experiments like the Electron Ion Collider \cite{kalicy2016high}.
In the following we will consider as an example the case of the \textsc{GlueX} experiment \cite{stevens2016gluex,patsyuk2018status} at the Jefferson Laboratory, where the DIRC has been recently installed utilizing components of the decommissioned BaBar DIRC to enhance the PID capabilities of the experiment.

\begin{figure*}
    \includegraphics[width=.65\textwidth]{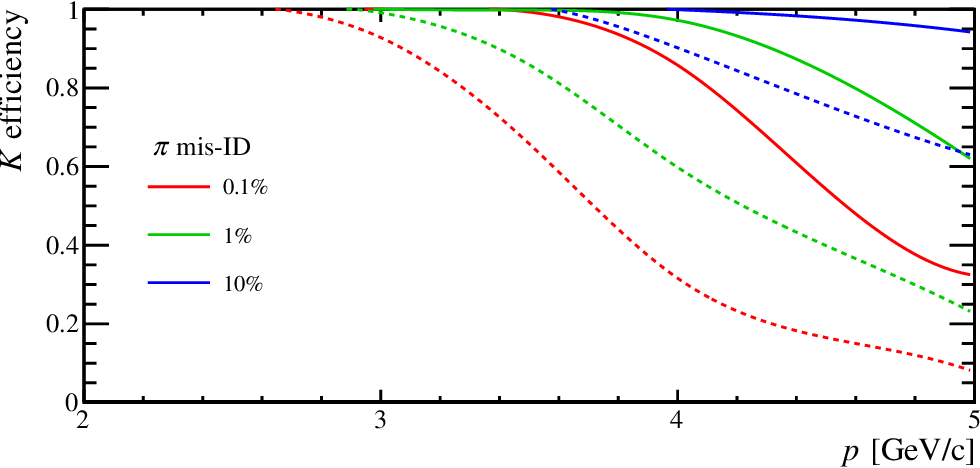}
    \caption{Kaon efficiency vs momentum for $\pi$ mis-identification probabilities of 0.1, 1, and 10\% (i.e. probability for a charged $\pi$ track to be incorrectly identified as a charged $K$). The dashed curves show the conservative performance, while the solid curves show the improved performance achieved in simulation for the current GlueX DIRC design. Image taken from \cite{stevens2016gluex}, where the reader can find more details.}
    \label{fig:expected_efficiency}
\end{figure*}

Our choice is motivated by FastDIRC \cite{hardin2016FastDIRC}, an open source simulation and reconstruction package for DIRC detectors implementing the \textsc{GlueX} DIRC geometry.
\begin{figure*}
    \includegraphics[width=.45\textwidth]{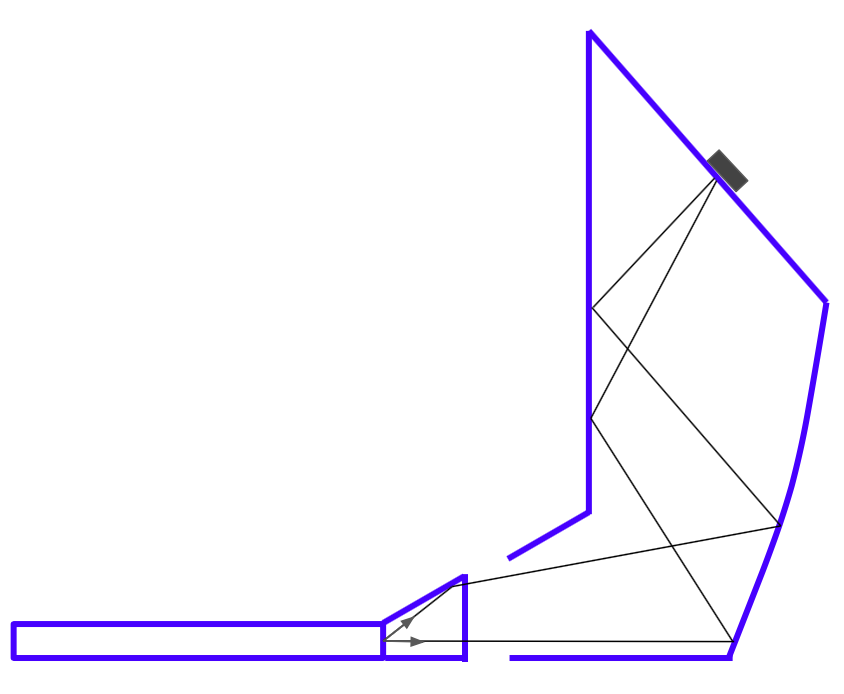}
    \includegraphics[width=.45\textwidth]{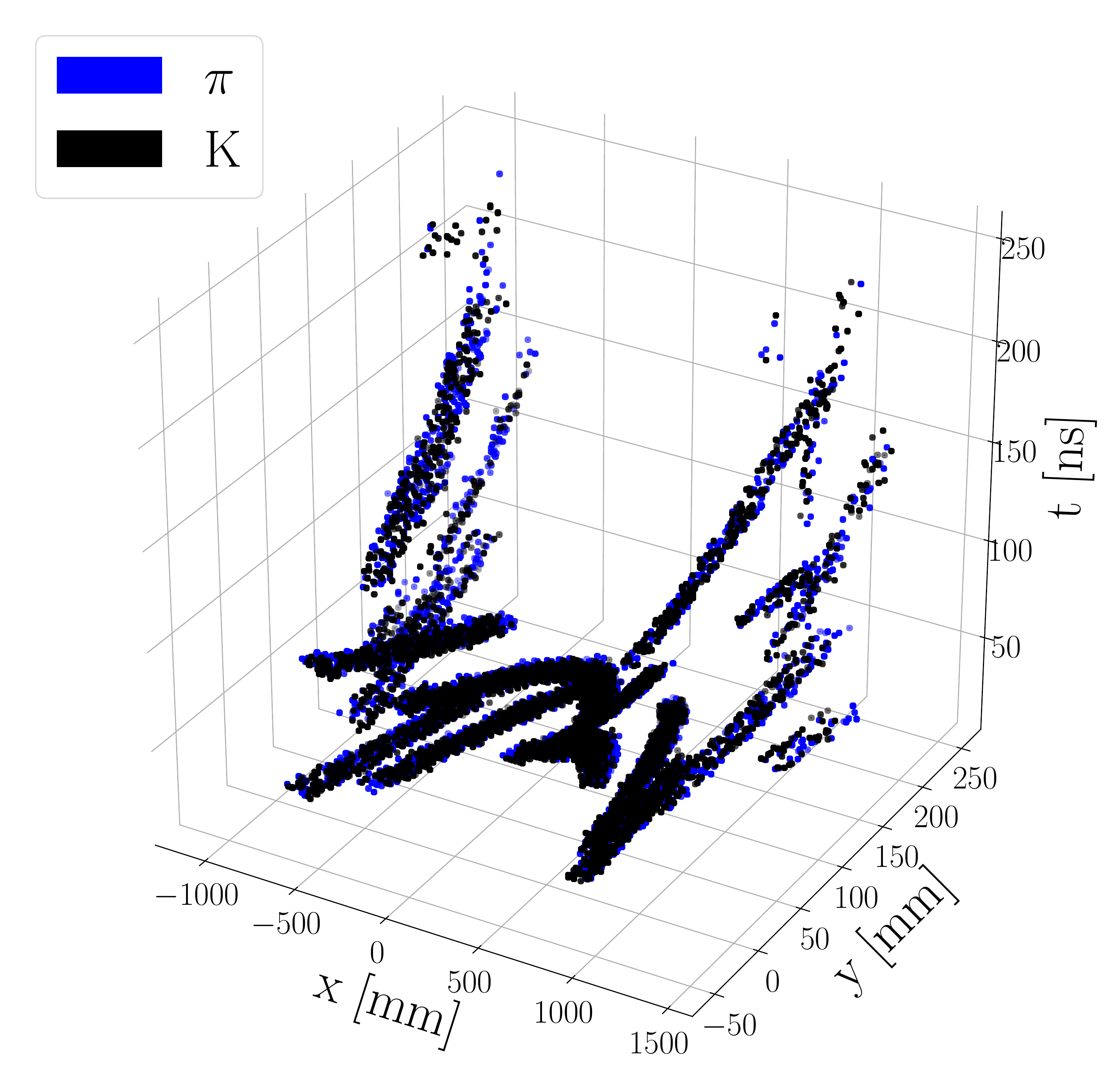}
    \caption{(left) Drawing (not to scale) of the \textsc{GlueX} DIRC box containing fused-silica bars and wedges. An optical box filled with distilled water is used to transport the Cherenkov photons from the wedges to the PMT arrays. (right) Example of hit pattern detected in the PMT plane (spatial coordinates are dubbed x,y, while the time is indicated as t) simulated with FastDIRC. The two colors correspond to the hit pattern of a kaon and of a pion as reported in the legend, under the same kinematic conditions (i.e. particle momentum, incident angle on the bar, azimuthal angle with respect to the bar, location on the bar and which bar has been hit).}
    \label{fig:sketch}
\end{figure*}
This geometry consists of four bar boxes and two photon cameras, where each bar box contains 12 fused silica radiators (1.725 $\times$ 3.5 $\times$ 490 cm$^{3}$). 
Both photon cameras are attached to two bar boxes and are equipped with an array of Multianode Photomultiplier Tubes (MaPMTs) allowing a three-dimensional (x,y,t) readout with a time resolution of approximately 200~ps. Patterns take up significant fractions of the PMT in x,y and are read out over 50-100~ns due to propagation time in the bars. The reader can find in Fig. \ref{fig:sketch} (left) a schematic of the detector with one of the two photon cameras and in Fig.~\ref{fig:sketch} (right) an example of hit pattern generated with FastDIRC expected in the PMT plane (x,y) as a function of the propagation time.
In particular, the \textsc{GlueX} experiment is designed to search for gluonic excitations in the meson spectrum produced through photoproduction reactions at a tagged photon beam facility. For this physics program, the DIRC is expected to provide a good separation power between pions and kaons of at least 3$\sigma$ up to 4 GeV/c in momentum (a plot of the kaon efficiency as a function of the kaon momentum for different pion mis-identification probabilities is shown in Fig. \ref{fig:expected_efficiency}), which allows systematic studies of kaon final states that are essential for inferring the quark flavor content of both hybrid and conventional mesons \cite{lacock1997hybrid,meyer2015hybrid}. 
For all these reasons, developing an efficient and fast reconstruction algorithm is of crucial importance. 
Notice that in the case of ring imaging Cherenkov (RICH) detectors, the time variable is typically not used in the reconstruction methods. This feature could be part of future reconstruction algorithms if better time resolutions are achieved.  
Instead in the DIRC case, the larger propagation times contribute to distinguish the type of particle producing Cherenkov light. 
Depending on the type of detector, DeepRICH reconstruction can be based on spatial features only or on combined space and time components. 

The outline of this paper is as follows: existing reconstruction methods are discussed in Sec. \ref{sec:other_methods}; the DeepRICH architecture is presented in Sec. \ref{sec:deep_RICH}; application to the DIRC case, discussion of the results and comparison with FastDIRC are described in Sec. \ref{sec:results}; summary and conclusions are reported in Sec. \ref{sec:summary}.

\section{Established Methods and Novel Approaches}\label{sec:other_methods}

Cherenkov detectors are relatively slow to simulate with full simulations like Geant \cite{GeantDIRC}---\textit{e.g.}, for the DIRC case, each Cherenkov photon reflects on average $\mathcal{O}(10^{2})$ times within a bar and this makes the simulation CPU intensive---thus new approaches are being developed to get a faster reconstruction of the detected light  \cite{hardin2016FastDIRC,dey2015design,derkach2019cherenkov,dzhygadlo2016panda}.
In this section we briefly describe the state of the art of established computational methods and provide an overview of novel paradigms based on machine learning.

\subsection{The Geometrical Reconstruction Method}

The geometrical reconstruction method is based on the BaBar DIRC algorithm \cite{dzhygadlo2014simulation}. 
This approach involves generating in advance a large number of photons at different angles exiting each bar, and then tracking them to the PMT plane.
In this way a look-up table is created, where each pixel on the photo-detection plane is associated to a set of photon directions at the exit from the bar potentially leading to a photon detected in that pixel. 
The Cherenkov angle $\theta_{C}$ of each photon is then reconstructed combining the particle direction provided by the tracking system with the photon direction taken from a look-up table.
The look-up table is stored as a ROOT tree with the size of about 300 MB \cite{patsyuk2018status}. 
The resulting cumulative distribution of the reconstructed Cherenkov angles is typically characterized by the peaks at the expected values of $\theta_{C}$ for pions and kaons and a combinatorial background beneath them. The width of the Cherenkov angle reflects the single photon Cherenkov angle resolution characteristic of the detector performance.

\subsection{Time-based Image Reconstruction}

Another approach is the so called time-based imaging reconstruction which is derived from a method used by the Belle II TOP \cite{starivc2008likelihood}.
For every particle hypothesis, the expected arrival time of Cherenkov photons is calculated analytically based on the charged particle direction and hit location and is compared to the measured time, yielding to likelihoods.  
This method is rather compute-intensive, as one in principle should simulate all the configurations of the charged particles as a function of the mass, energy, direction and location in the DIRC bars.

\subsection{FastDIRC}

The main characteristic of the FastDIRC algorithm \cite{hardin2016FastDIRC} is to analytically trace the photons through the optical system. This approach is about $\mathcal{O}(10^{4})$ times faster than the full Geant simulation. 
The reconstruction is based on a kernel density estimation (KDE) \cite{cranmer2001kernel} of the probability distribution function (PDF) for each assumed particle type. 
The expected distributions on the detection plane for each charged particle hypothesis are compared to the actually observed hit patterns to build likelihoods.
FastDIRC allows for parameterization, a feature that makes it suitable for detector design optimization and for offline calibration of real data. 
It has been shown \cite{hardin2016FastDIRC} that the resolution of the reconstructed Cherenkov angle is about 30\% better than the geometric reconstruction method. 
However the FastDIRC method is about $\mathcal{O}(10^{2}-10^{3})$ times slower than the look-up table based reconstruction.

In this paper, FastDIRC is used as a source of reliable simulated events that are injected as input of the DeepRICH architecture.

\subsection{Generative Adversarial Network}

A first attempt to apply deep learning to simulate Cherenkov detector response appeared recently in \cite{derkach2019cherenkov}, where it has been proposed to use a generative adversarial neural network (GAN) \cite{goodfellow2014generative} to bypass low-level details at the photon generation stage. 
This work is based on events simulated with FastDIRC assuming the design of the \textsc{GlueX} DIRC.
The GAN architecture is trained to reproduce high-level features (the likelihood results from FastDIRC) based on input observables of the incident charged particles, allowing for an improvement in simulation speed. 
The authors of \cite{derkach2019cherenkov} claim a good precision and very fast performance (the batch generation on GPU produces up to 1 million track predictions per second) from their studies. 
\\Recently in another paper \cite{maevskiy2019fast} generative models have been used for fast simulation of RICH detectors at LHCb.

In the following section we are going to present a new deep architecture called DeepRICH, providing a thorough description of the code, data preparation, training/testing phases and performance.

\section{The DeepRICH Network}\label{sec:deep_RICH}

Differently from the GAN based method, which directly maps the injected input to the reconstructed output, our generative model explicitly reconstructs the injected hit patterns expected for each kinematics, and internally creates latent variables that allow to classify the particles.  

\subsection{Architecture}

DeepRICH is based on a custom Variational Auto-encoder \cite{VAE}.
VAEs are generative models that try to simulate how the data are generated. 
In order to characterize the causal relations underlying the observed data, VAEs provide a posterior function approximated by an autoencoder architecture, which is made by an encoder and a decoder, the latter being symmetric to the first in terms of layer structure.

In what follows we describe each detected hit by a three-dimensional vector, (x,y,t), corresponding to the %
spatial and temporal components. 
We use the notation $\mathbf{x} \in \mathbb{R}^{m \times 3}$ to indicate $m$ hits associated to an individual charged particle. The kinematic parameters of each particle are represented by the vector $\mathbf{h}$, and they embody information on the particle momentum, angle and location where the particle crossed each bar (more details on this can be found in Sec.~\ref{subsec:data}, where we discuss about the preparation of data). 

\begin{figure*}
\centering
\includegraphics[width=.57\textwidth]{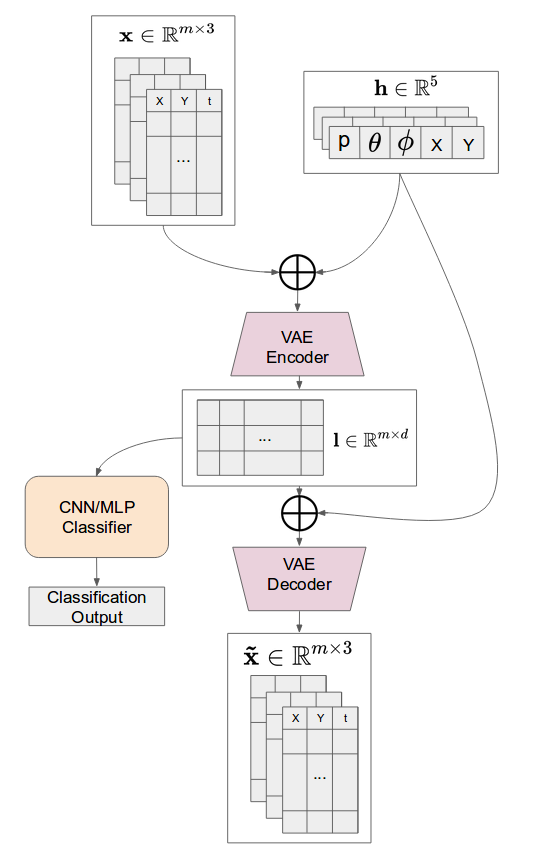}
\caption{A flowchart of DeepRICH: the inputs are concatenated---n.b., the $\oplus$ represents the concatenation between vectors---and fed into the encoder, which generates a set of vectors of latent variables, which are then used for both the classification of the particle and for the reconstruction of the hits.}
\label{fig:model}
\end{figure*}

Our novel architecture consists of three main parts:

\begin{itemize}
    \item An \textbf{Encoder}, which takes as input the concatenation between (i) $m$ hits produced by a particle, $\mathbf{x} \in \mathbb{R}^{m \times 3}$ and (ii) the associated vector $\mathbf{h}$ of kinematic parameters, to produce a $d$-dimensional vector of latent variables for each input hit, i.e. $\mathbf{l}\in \mathbb{R}^{m\times d}$.
    \\These vectors contain all the information that the network is capable of extracting from the hits $\mathbf{x}$.
    
    \item A \textbf{Decoder}, which takes as input the vectors of latent variables $\mathbf{l}$ concatenated with $\mathbf{h}$ and provides as output a set of hits $\mathbf{\tilde{x}} \in \mathbb{R}^{m \times 3}$, corresponding to the reconstruction of the input $\mathbf{x}$.
    
    \item A \textbf{Particle Classifier}, which basically consists in convolutional and linear layers; the network takes as input the vectors of latent variables $\mathbf{l}$ to classify the particle. 
    \\The challenging aspect here is to use the information extracted from the Encoder
    to do PID, that is to understand if the particle that has generated the hits %
     $\mathbf{x}\in \mathbb{R}^{m \times 3}$
    is a pion ($\pi$) or a kaon ($K$).\footnote{We are focused here on distinguishing $\pi$s from $K$s as this is the main scope of the \textsc{GlueX} DIRC. DeepRICH can be generalized to more than two categories of particles.}
    
\end{itemize}
A flowchart of the DeepRICH network is represented in Fig.~\ref{fig:model}.

The model is trained by minimizing the total loss function which is:
\be \label{loss_tot}
\begin{multlined}
\mathcal{L}(\mathbf{x}, \mathbf{\tilde{x}}, \mathbf{y}, \mathbf{\tilde{y}}, \mathbf{l}) = \\
\lambda_r \mathcal{L}_r(\mathbf{x}, \mathbf{\tilde{x}}) + \lambda_c \mathcal{L}_c(\mathbf{y}, \mathbf{\tilde{y}}) + \lambda_v \mathcal{L}_v(\mathbf{l}), 
\end{multlined}
\ee
where the $\lambda$ multipliers are used to weigh the contribution of the corresponding loss terms, described in the following:

(i) The term $\mathcal{L}_r$ is the average reconstruction loss between the real particle $\mathbf{x}$ and the output of the Decoder $\mathbf{\tilde{x}}$, calculated using the L1 smooth loss (also called Huber error. See, \textit{e.g.},~\cite{Girshick_2015}):\footnote{Notice we are using a simplified notation: another sum running over all the hits of the particle is present in Eq.~\eqref{loss_r}.}
\be \label{loss_r}
\mathcal{L}_r(x, \tilde{x}) = \frac{1}{3}\sum_{i}^{3} z_i
\ee 
where $z_i$ is given by
\[
    z_i= 
\begin{dcases}
    0.5(x_i - \tilde{x}_i)^2,& \text{if } |x_i - \tilde{x}_i| < 1\\
    |x_i - \tilde{x}_i| - 0.5              & \text{otherwise,}
\end{dcases}
\]
and the index $i$ indicates the spatial or time components of each hit.
Such a loss is less sensitive to outliers than the Mean Squared Error (MSE). 
\\In fact in the case of an unbounded output, MSE requires careful tuning of the learning rate and the loss in order to prevent exploding gradients.

(ii) The term $\mathcal{L}_c$ is the classification accuracy, calculated using the Cross Entropy between the target $y$, i.e. the ground truth particle's type (0 for kaons and 1 for pions), and the output of the classification layer $\tilde{y}$. 
\be \label{loss_c}
\mathcal{L}_c = -(\mathbf{y} \log (\mathbf{\tilde{y}}_0) + (1- \mathbf{y}) \log(\mathbf{\tilde{y}}_1))
\ee 
where the components $\mathbf{\tilde{y}_0}$ and $\mathbf{\tilde{y}_1}$ refer to the logits, associated to pions and kaons, scaled using the $softmax(\cdot)$ function. After this scale we have $\mathbf{\tilde{y}_0} + \mathbf{\tilde{y}_1} = 1$.

(iii) The loss $\mathcal{L}_v$ is a term calculated using the Maximum Mean Discrepancy (MMD) \cite{MMD}, as explained in the following; notice that the idea of combining VAE and MMD was used for the first time in \cite{infoVAE}, 
where the authors proved that infoVAE (VAE using MMD) is fast to train, stable and leads to a better learning of the features if compared to the traditional evidence lower bound (ELBO) \cite{hoffman2016elbo} criterion used in VAEs. 
The basic idea of MMD is that two distributions are identical if and only if their moments are the same. 
Assuming to have two distributions $\textbf{\textit{p}}(z)$ and $\textbf{\textit{q}}(z)$, one can measure the divergence between these distributions:
\be \label{loss_v}
\begin{multlined}
\mathcal{L}_v = MMD(\textbf{\textit{p}}(z), \textbf{\textit{q}}(z)) =   
\mathbb{E}_{\textbf{\textit{p}}(z), \textbf{\textit{p}}(z')}[\kappa(z, z')]  + \\ 
\mathbb{E}_{\textbf{\textit{q}}(z), \textbf{\textit{q}}(z')}[\kappa(z, z')] 
-2 \mathbb{E}_{\textbf{\textit{p}}(z), \textbf{\textit{q}}(z')}[\kappa(z, z')],
\end{multlined}
\ee
where $\kappa ( \cdot , \cdot )$ can be any positive definite kernel, which can be seen as a function that measures the distance between two samples. To this end, we use a Gaussian kernel \cite{scholkopf2001kernel}. In our case the distribution $\textbf{\textit{p}}(z)$ is related to the vector of latent variables, and $\textbf{\textit{q}}(z)$ is a normal distribution $ \mathcal{N}(0, \sigma)$; the best value of $\sigma$ is determined using the Bayesian optimization described in Sec. \ref{sec:optimization}. A naive intuition of MMD is that the latent vectors should follow the same distribution of $\textbf{\textit{q}}(z)$.

The architecture described in this section is also summarized in form of a pseudo-code in the Alg. \ref{pseudocode}.

\begin{figure}
\begin{minipage}{\linewidth}
\begin{algorithm}[H]
\caption{Pseudo-code for particle identification with DeepRICH}
\label{pseudocode}
\begin{algorithmic}[0]
\Procedure{training}{}
\For {$i=1 \dots \text{E}$ (training epochs)}
\For {each batch $b=(\mathbf{x}, \mathbf{y}, \mathbf{h}) \in \text{B}$}\\
\State {$\mathbf{l} \leftarrow Encoder(\mathbf{x}, \mathbf{h}; \ \theta)$} 
\State {$\mathbf{\tilde{y}} \leftarrow Classifier(\mathbf{l}; \ \theta)$} 
\State {$\mathbf{\tilde{x}} \leftarrow Decoder(\mathbf{l}, \mathbf{h};\ \theta)$}
\State {update $\theta$ by minimizing total loss $\mathcal{L}(\mathbf{x}, \mathbf{\tilde{x}}, \mathbf{y}, \mathbf{\tilde{y}}, \mathbf{l})$}\\
\EndFor
\EndFor
\EndProcedure
\vspace{0.25cm}
    \begin{itemize}
        \item \small{$\mathbf{x} \in \mathbb{R}^{m \times 3}$ is a set of hit produced by a charged particle; $\mathbf{y}$ is the ground truth of the particle (i.e., a $\pi$ or a $K$); $\mathbf{h}$ is the vector containing the kinematic parameters associated to the particle; $\theta$ are the weights (parameters) of the networks; $\mathbf{l}$ is the vector of latent variables associated to $\mathbf{x}$ and produced by the Encoder.}
        \item \small{For each hit in the particle, the Encoder produces a vector of the latent variables  $\mathbf{l}$, by taking as input the encoded kinematic parameters concatenated with the hit itself.}
        \item \small{The vectors of latent variables associated to the hits of a particle are used to classify the particle itself.}
        \item \small{The Decoder reconstructs the input hits using the latent variables and the kinematic parameters.}
    \end{itemize}
\end{algorithmic}
\end{algorithm}
\end{minipage}
\end{figure}

In addition we use a dropout layer after each layer in the Encoder/Decoder, with drop probability equal to 10\%; we also apply a dropout on the latent variables before feeding them into the CNN, with a probability equal to 50\%. 
We fix the number of layers in the decoder/encoder to 2, while the number of hidden unites is set to [512, 256]. 
The CNN has 3 layers with, respectively, [64, 64, 128] kernels with stride 1 and size 3, whereas the classifier has 4 layers with [100, 50, 25, 2] neurons, where the dimension of the last layer correspond to the number of classes ($\pi$ and $K$). %
The activation function used after each layer is the Rectified Linear Unit (ReLU). The reader can find more technical details summarized in Table~\ref{tab:network}.

\begin{table*}[]
\resizebox{0.9\textwidth}{!}{%
\begin{tabular}{|c|c|c|c|c|}
\hline
\multicolumn{1}{|c|}{\textbf{Architecture name}} & \textbf{Type} & \multicolumn{1}{l|}{\textbf{Neurons/kernel size}} & \textbf{Activation function} & \textbf{Regularization} \\ \hline
\multirow{3}{*}{Encoder/Decoder} & \multirow{3}{*}{Linear} & 512/256 & ReLU & Drop 0.1 \\ \cline{3-5} 
 &  & 256/512 & ReLU & Drop 0.1 \\ \cline{3-5} 
 &  & Latent Dim/3 &  &  \\ \hline \hline
\multirow{3}{*}{CNN} & \multirow{3}{*}{Conv} & 64 (3x3, Stride 1) & ReLU & Batch Norm \\ \cline{3-5} 
 &  & 64 (3x3, Stride 1) & ReLU & Batch Norm \\ \cline{3-5} 
 &  & 128 (3x3, Stride 1) &  & Batch Norm + Drop 0.5 \\ \hline \hline
\multirow{4}{*}{Classifier} & \multirow{4}{*}{Linear} & 100 & ReLU &  \\ \cline{3-5} 
 &  & 50 & ReLU &  \\ \cline{3-5} 
 &  & 25 & ReLU &  \\ \cline{3-5} 
 &  & 2 &  &  \\ \hline
\end{tabular}%
}
\caption{A detailed description of the DeepRICH architecture used in the experiments. Each sub-architecture---specified by the type of neural network---is described layer by layer in terms of the number of neurons and the size of the kernel, the used activation function and the regularization. }
\label{tab:network}
\end{table*}

\subsection{Data Preparation}\label{subsec:data}
The data generation is based on FastDIRC \cite{hardin2016FastDIRC}. 
FastDIRC allows to generate the hit pattern observed in the PMT detection plane for a given kinematics of the charged particle traversing the radiator. 
The kinematics is characterized by different parameters, namely the momentum of the particle $p$ [GeV/c], the polar angle $\theta$ relative to the normal to the surface of the bars, the azimuthal angle $\phi$, the location $X$, $Y$ on the surface of the bar, the information (as an integer index) on which fused silica bar has been hit.\footnote{We use capital letters to distinguish the location on the bar ($X$,$Y$) from the hit spatial coordinates ($x$,$y$) in the detection plane.}  
FastDIRC use kernel density estimation to produce an estimate of the probability distribution function on the PMT plane. It generates about $10^{5}$ provisional points for each kinematics, which are used to detect an actual charged particle passing through the bars and generating a sparse hit pattern of about 20-50 ``real'' hits.  
FastDIRC therefore generates both the sparse hit patterns associated to one particle as well as the whole probability density function (PDF) associated to a particular kinematics which is used to identify that particle. 
The training set for DeepRICH has been generated with FastDIRC combining more than one kinematics for a single bar. 
A particular region of the phase-space can be divided into a fine grid of points.
For example, the largest dataset we generate corresponds to an hypercube of $\sim$ 2$\cdot$10$^{4}$ kinematic points covering the kinematic subspace \textup{$\Delta p \times \Delta\theta \times \Delta\phi \times \Delta X \times \Delta Y$ = [4, 5] [GeV/c] $\times$ [2, 5] [deg] $\times$ [20, 90] [deg] $\times$ [-17.5, 17.5] [mm] $\times$ [50, 1000] [mm]}, where $\theta$, $\phi$, $X$ and $Y$ have been divided into equally distant points within those intervals.
For each point of the grid we generate one PDF with FastDIRC and then sample randomly the observed ``real'' hits. This is done by taking into account the expected yield of the photons: we implemented an yield generation inspired by the FastDIRC simulation of the observed hits which takes into account the photon yield reduction due to several effects, \text{e.g.}, if the total internal reflection condition is not met or a photon misses a mirror.
We also check that keeping the yield constant (fixing it to 40 photons) does not change the performance significantly. 

Consistently with the expectations, a more dense grid of points combined with a larger number of sampled particles at each kinematic point generally improves the PID performance of DeepRICH (this can be quantified as the Area Under Curve described in Sec.~\ref{sec:comparison_fastdirc}). 
Taking into account that the intrinsic limit on the achieved performance depends on the kinematic conditions (\textit{e.g.}, the larger the momentum the lower is the $\pi/K$ distinguishing power), a tradeoff on the above numbers (i.e. how dense the grid and how many particles should be chosen for training) can be found based on the sought classification accuracy and the available computing resources. 

\subsection{Model Training and Testing } 

At each kinematic point ($p, \theta, \phi, X, Y$) we use FastDIRC to produce a large number of expected hits for both $\pi$s and $K$s. 
Then we sample $N$ particles of a given type ($\pi$ or $K$) where 
by construction each particle consists of a random set of $m$ hits. %
In this way we avoid that the network learns how to classify particles based on some patterns internal to the FastDIRC generation algorithm. 
At the same time with this choice we can virtually build an unlimited dataset of particles from the PDFs of FastDIRC. 

The generated samples have been then divided into two subsets: training and test: 
(i) The training set contains particles at certain kinematics which are used during the training phase---ensuring that all the vertices of the hypercube are included---while (ii) the test subset will be used only for testing the network performance after the training procedure to see if it can achieve good results on unknown kinematics. 
Furthermore the particles from the training set are divided into ``training particles'' and ``development particles'' (the split is 80\%/20\%); the training particles are used to update the parameters of the network by minimizing the total loss (see Eq.~\eqref{loss_tot}), while the development particles are used to calculate an accuracy score, to evaluate the goodness of the classification while training and check if the network is learning properly how to classify hits from known kinematics.
Early stopping is used to interrupt the training procedure if the development score does not improve after a certain number of epochs. The classification score on the development particles is also used to tune the hyperparameters of the network with a Bayesian optimization (the procedure is explained in detail in Sec. \ref{sec:optimization}). 

We then optimize the parameters of the network with Adam \cite{Kingma2014AdamAM} using the tuned learning rate. %
 The dataset has been standardized---for each feature we choose 0 mean and standard deviation (Std) equal to 1---and this is done separately for both the hits and the kinematics parameters, in order to avoid the overshadowing of features with smaller values and further improve the training procedure; notice that the development and test hits have been standardized using the mean and the Std calculated on the training hits, to avoid a potential injection of bias that could improve the classification performance.

We train the network in different experiments, each consisting of at most 50 epochs, and evaluate the performance on the development subset during the training phase. 
The development accuracy is calculated by applying the $softmax(\cdot)$ on the classification layer. 
When the training is over, the model is evaluated on the test particles extracted from unknown kinematics. 

\subsection{Network Optimization} 
\label{sec:optimization}

Bayesian Optimizers (BOs) \cite{jones1998efficient, snoek2012practical} are among the most efficient tools for optimizing the hyperparameters of a deep architecture \cite{eggensperger2013towards}.
In fact BOs search for the global optimum $x^{*}$ over a bounded domain $\chi$ of a black-box functions $f(x)$.
In particular, $f$ can be noisy, non-differentiable and expensive to evaluate. 

Typically gaussian processes \cite{williams2006gaussian} are used to build a surrogate model of $f$, but other regression methods such as decision trees can also be used. 
Once the probabilistic model is determined, a cheap utility function (also called acquisition function) is considered to guide the process of sampling the next point to evaluate. 
The DeepRICH network consists of N hyperparameters listed in Table~\ref{tab:hyper_settings}. In particular, the multipliers of the loss functions defined in Eqs. \eqref{loss_r}, \eqref{loss_c}, \eqref{loss_v}, the dimension of the latent variables, the MMD variance and the learning rate play an important role in the performance of the network.  
These hyperparameters are tuned with a BO provided by the sklearn \cite{skopt} package.
As previously discussed, other hyperparameters, \textit{e.g.}, the number of layers in the architecture, are not tuned and their values are reported in Table~\ref{tab:network}.
We choose as objective function $f$ the development score obtained during the training phase. 
Each call of the BO is based on 50 epochs. %
Results of the optimization are summarized in Table~\ref{tab:hyper_settings}. 

\begin{table*}
\caption{
\label{tab:hyper_settings}
List of hyperparameters tuned by the BO. The tuned values are shown in the outermost right column. The optimized test score is about 92\%.} 

\centering
\scalebox{1.}{
        \begin{tabular}{ | c | c | c | c |}
        \hline
        symbol & description & range & optimal value \\
        \hline
        \texttt{NLL} & $\lambda_r$ & [10$^{-1}$,10$^{2}$] & 0.784\\
        \texttt{CE} & $\lambda_c$ & [10$^{-1}$,10] & 1.403\\
        \texttt{MMD} & $\lambda_v$ & [1,10$^{3}$] & 1.009 \\
        \texttt{LATENT\_DIM} & latent variables dimension & [10,200] & 16\\
        \texttt{var\_MMD} & $\sigma$ in $\mathcal{N}(0, \sigma)$ & [0.01,2] & 0.646\\
        \texttt{Learning Rate} & learning rate & [0.0001, 1] & 6.6$\cdot$10$^{-4}$ \\
        \hline
    \end{tabular}
}
\end{table*}

\section{Results}\label{sec:results}

\begin{figure*}
    \includegraphics[height=.45\textwidth]{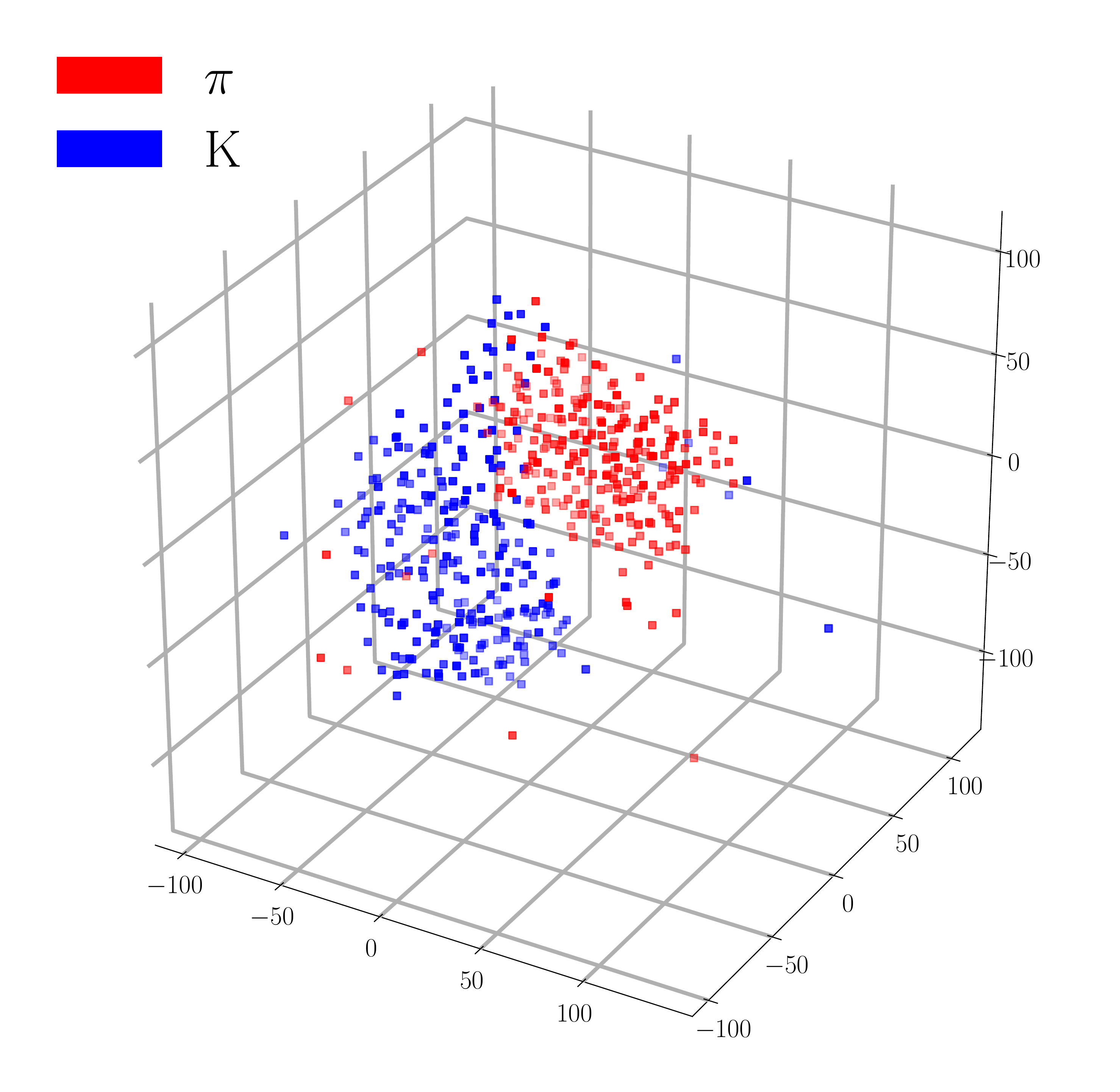}
    \includegraphics[height=.45\textwidth]{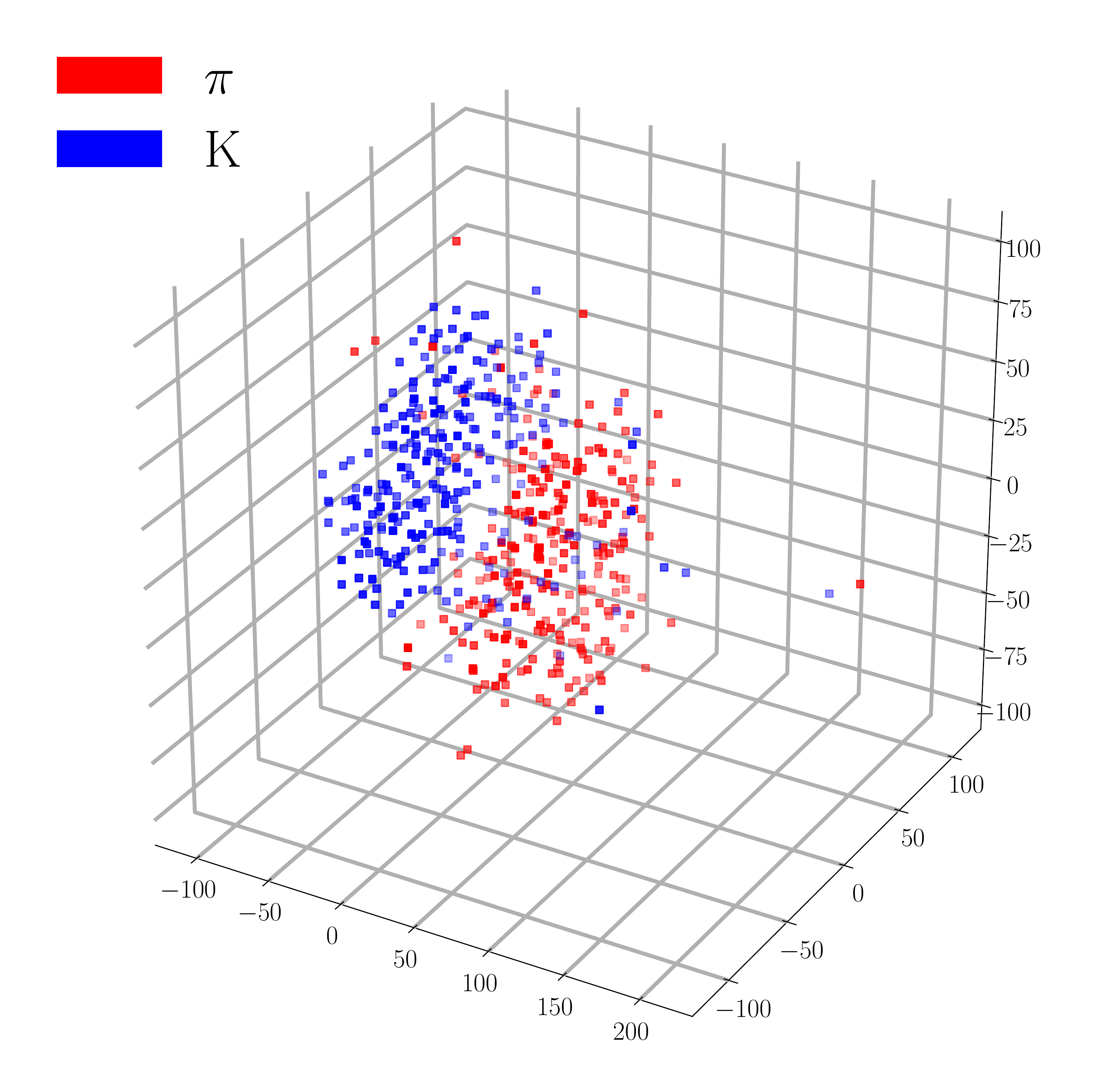}
    \caption{Example of features extracted by the CNN module from $\pi$'s and $K$'s at 4 GeV/c (left) and 5 GeV/c (right).
    These features are then used to classify the particle. The plot shows a better separation between $\pi$/$K$ at 4 GeV/c, which means that the network has good distinguishing power. As expected the points become less separated at larger momentum. The 3D visualization is obtained with t-SNE~\cite{maaten2008visualizing}.}
    \label{fig:pattern1}
\end{figure*}

The following results are based on charged $\pi,K$ candidates with momentum between 4 and 5 GeV/c, the latter corresponding to a challenging kinematics given the sizeable overlap between the expected hit patterns. 
The capability of distinguishing $\pi$s from $K$s and effectively doing PID depends on the features and the causal relations learnt in the space of the latent variables. 
A 3D visualization in the space of the latent variables is shown in Fig. \ref{fig:pattern1}, where t-SNE \cite{maaten2008visualizing} is used for dimensionality reduction. A clearer separation is achieved in the reduced space of the latent variables at 4 GeV/c compared to 5 GeV/c.

\begin{figure*}
    \centering
    \includegraphics[height=0.65\textwidth]{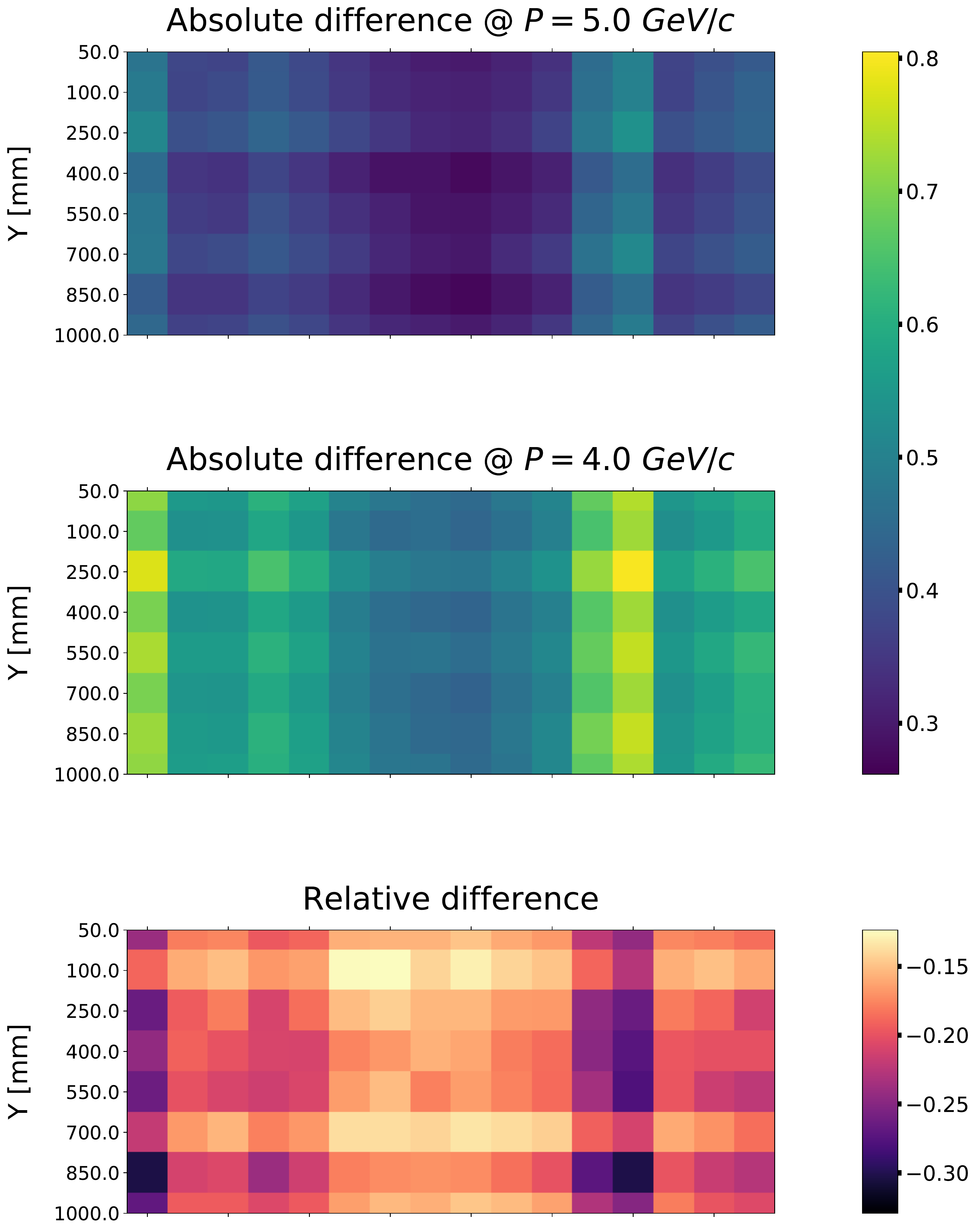}
    \caption{(Top and middle plots) 2D plot of the absolute difference on each latent variable between $\pi$'s and $K$'s, obtained for 5 GeV/c and 4 GeV/c, respectively. 
    The latent variables are binned along the x-axis, whereas the y-axis shows where the particle traversed the bar (along Y [mm]). The color axis indicates the value of the absolute difference. The larger the difference the larger is the distinguishing power of the network. As expected the separation becomes less clear when the momentum is larger whereas there is no appreciable dependence on the position on the bar resulting in patterns with vertical bands. (Bottom) The relative difference between the first row (5 GeV/c) and the second row (4 GeV/c) showing negative values in the majority of the bins. %
    } 
    \label{fig:pattern2}
\end{figure*}

An alternative representation of the same data is shown in Fig.~\ref{fig:pattern2}. 
Here the distinguishing power is quantified as the average absolute difference between $\pi$ and $K$ in each latent variable versus the Y-position on the quartz bar. 
This is shown at 5 and 4 GeV/c in momentum (top and middle of Fig.~\ref{fig:pattern2}, respectively). Notice that the number of bins (16 on the x-axis) corresponds to the dimension of the vector of latent variables. 
Intuitively, the larger the absolute difference the more $\pi$s are separated from $K$s. 
The relative difference (bottom of Fig.~\ref{fig:pattern2}) is characterized by negative values only, pointing to the obvious interpretation that the distinguishing power is larger at 4 GeV/c.
Notice also that in good approximation the separation between the two particle types does not depend on the Y-location on the quartz bar and we verify as a sanity check the presence of vertical bars in the patterns of Fig.~\ref{fig:pattern2} along the y-axis.

As described in Sec. \ref{subsec:data}, the event generation is based on FastDIRC which is also used in this section as a reconstruction algorithm to provide a benchmark against which evaluating the performance of the DeepRICH architecture.  

\subsection{Comparison with FastDIRC}\label{sec:comparison_fastdirc}

The PID strategy in FastDIRC is likelihood-based: $N_{d}$ photons for each candidate particle are detected in the PMT plane, and $N_{g}$ photons are generated to produce the expected PDFs of the 2 candidates ($\pi$, $K$). The $N_{d}$ particles are then used to compute the log-likelihood from each candidate PDF as follows:
\be \label{eq:logL}
\log{\mathcal{L}}_{\pi(K)} = \sum_{j=1}^{N_{d}}\ln{(\sum_{i=1}^{N_{g}^{\pi (K)}}g(\frac{|\mathbf{x}_{i}^{\pi (K)}-\mathbf{x}_{j}|}{\lambda})),}
\ee
where $\lambda$ is a bandwidth and $\mathbf{x}$ is a vector whose components are the spatial and time coordinates of each hit (either detected or generated).\footnote{In FastDIRC $N_{g}$ is chosen such that the achieved resolution reaches a stable value, and the bandwidth is tuned to provide the best performance.}   

The operational definition of likelihood in DeepRICH is different from Eq. \eqref{eq:logL}, in that different quantities are provided by the network: as explained in Sec.~\ref{sec:deep_RICH}, the output of the classifier is a two-dimensional vector $\mathbf{\tilde{y}} \in \mathbb{R}^{2}$, and we use these values as likelihoods for $\pi$ and $K$.

\begin{figure*}[!]
    \centering
    \includegraphics[width=.35\textwidth,height=.35\textwidth]{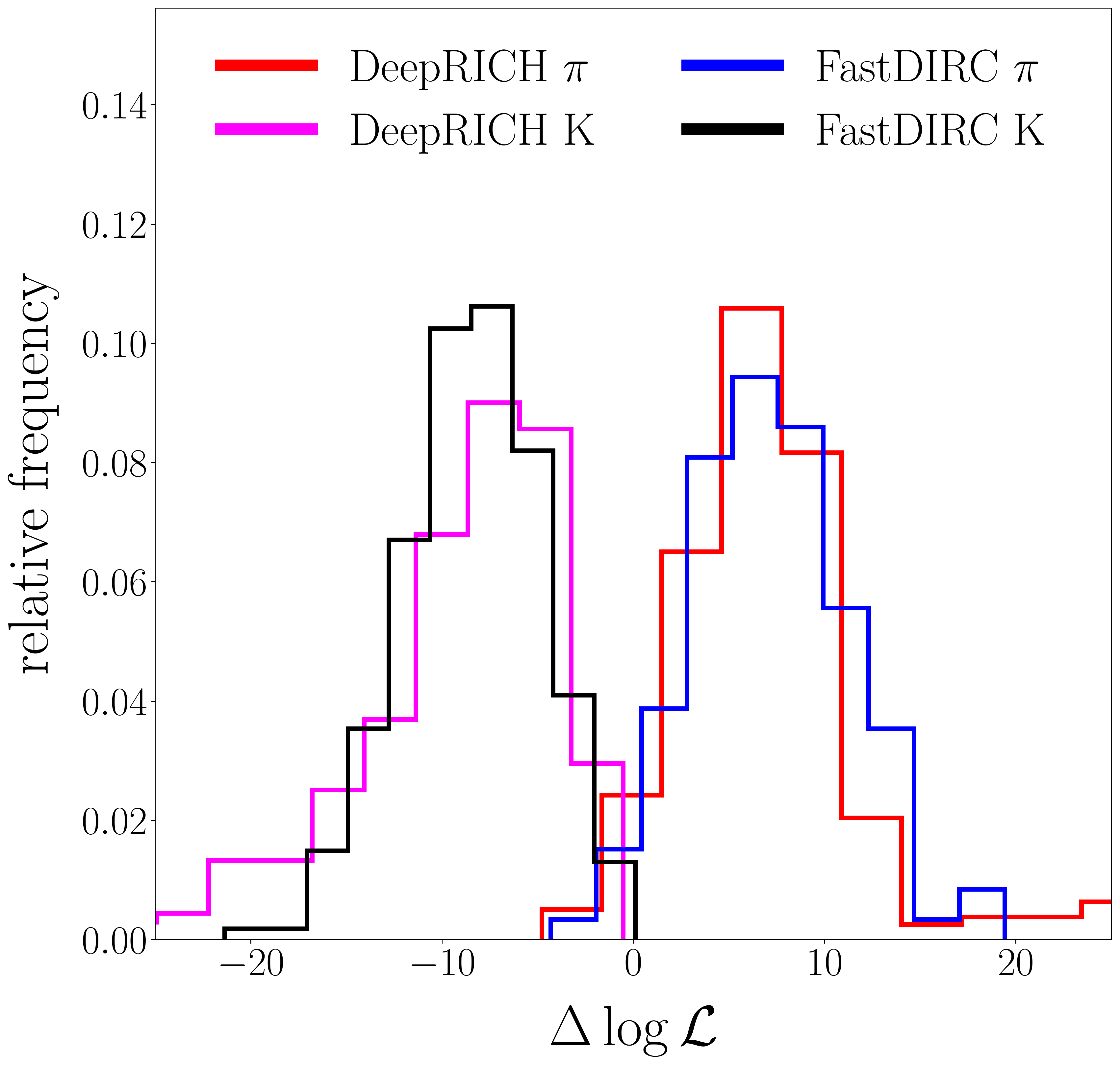}
    \includegraphics[width=.35\textwidth,height=.35\textwidth]{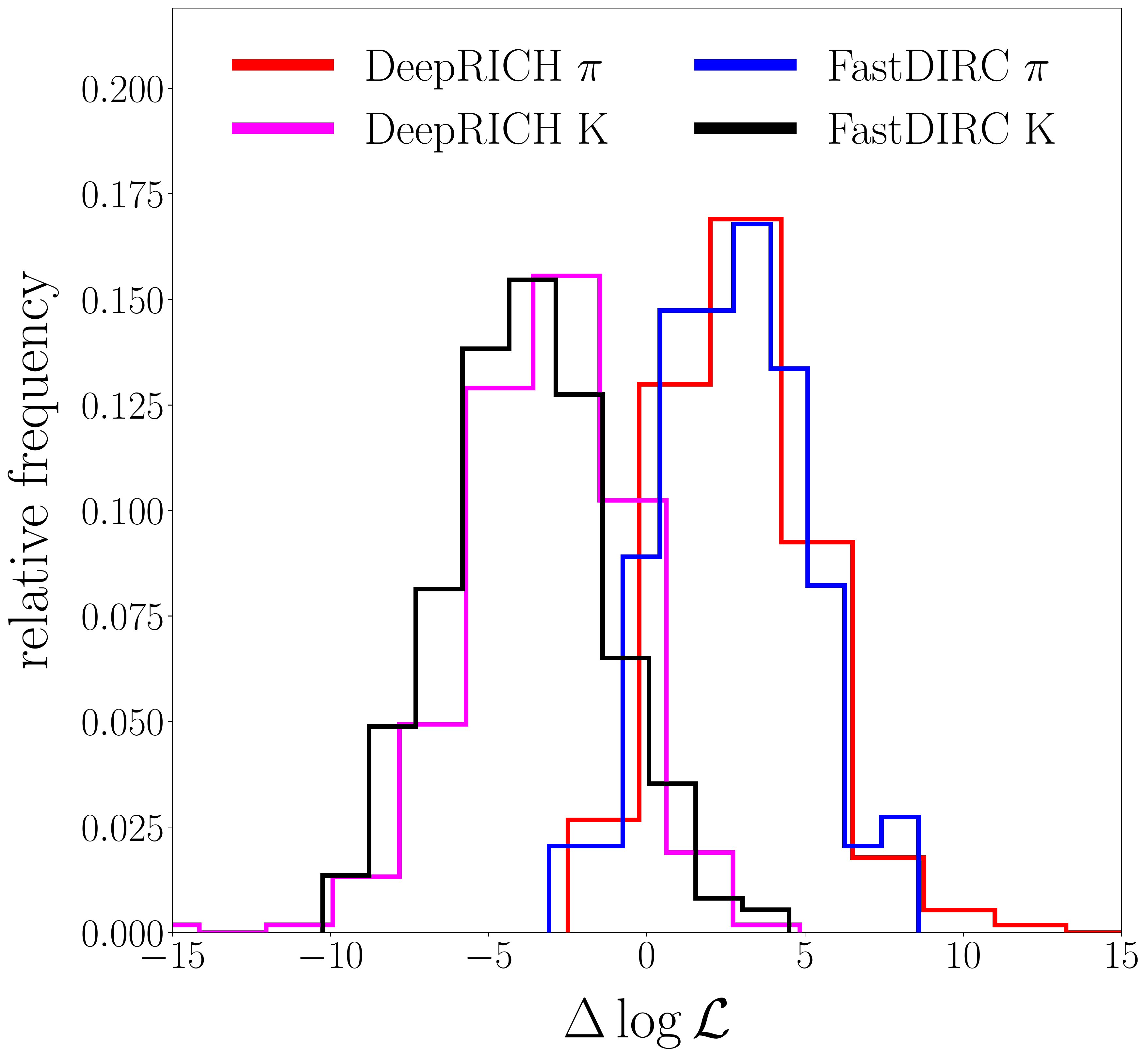}
    \\
    \includegraphics[width=.35\textwidth,height=.35\textwidth]{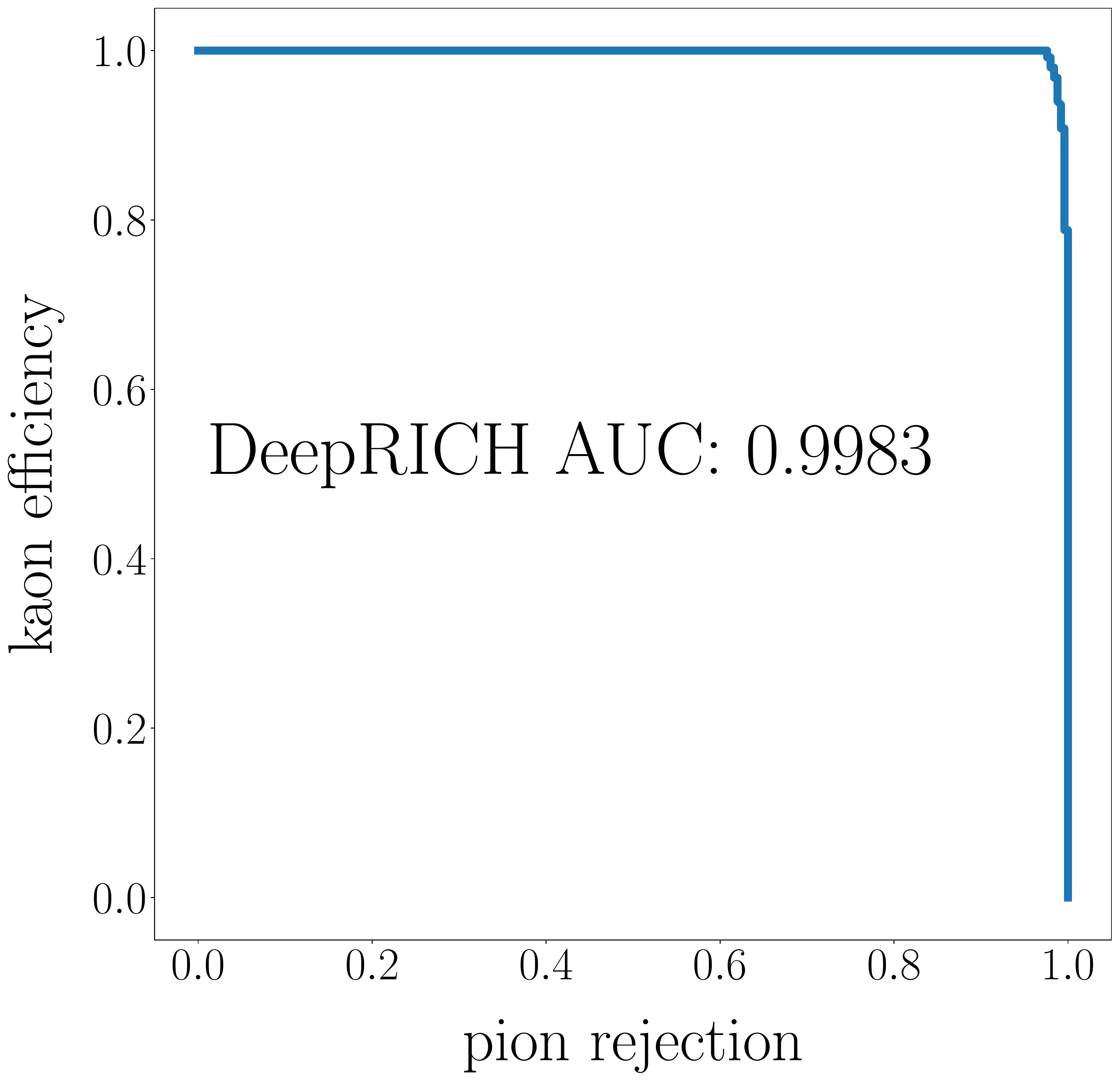}
    \includegraphics[width=.35\textwidth,height=.35\textwidth]{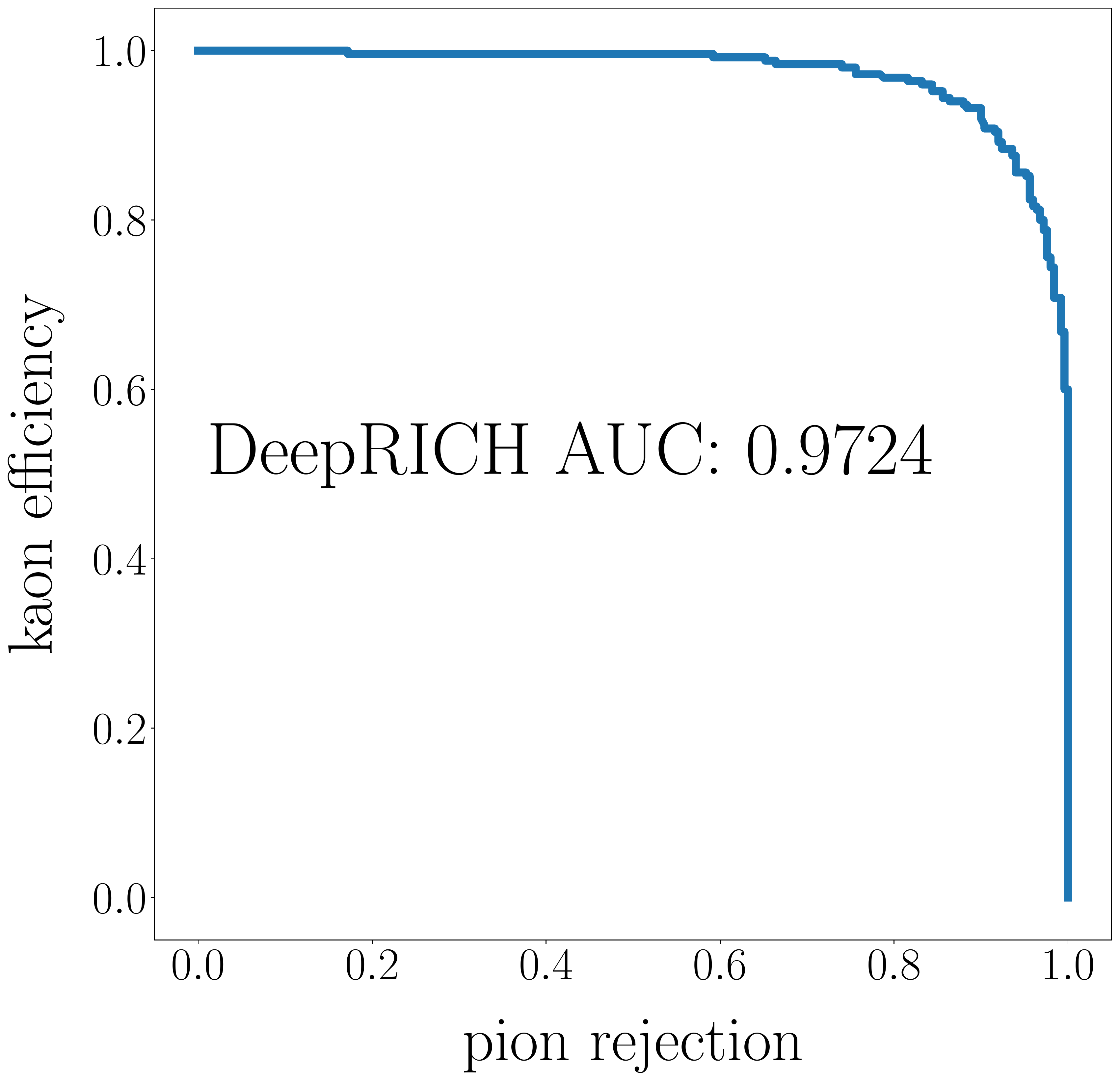}\\
    \includegraphics[width=.35\textwidth,height=.35\textwidth]{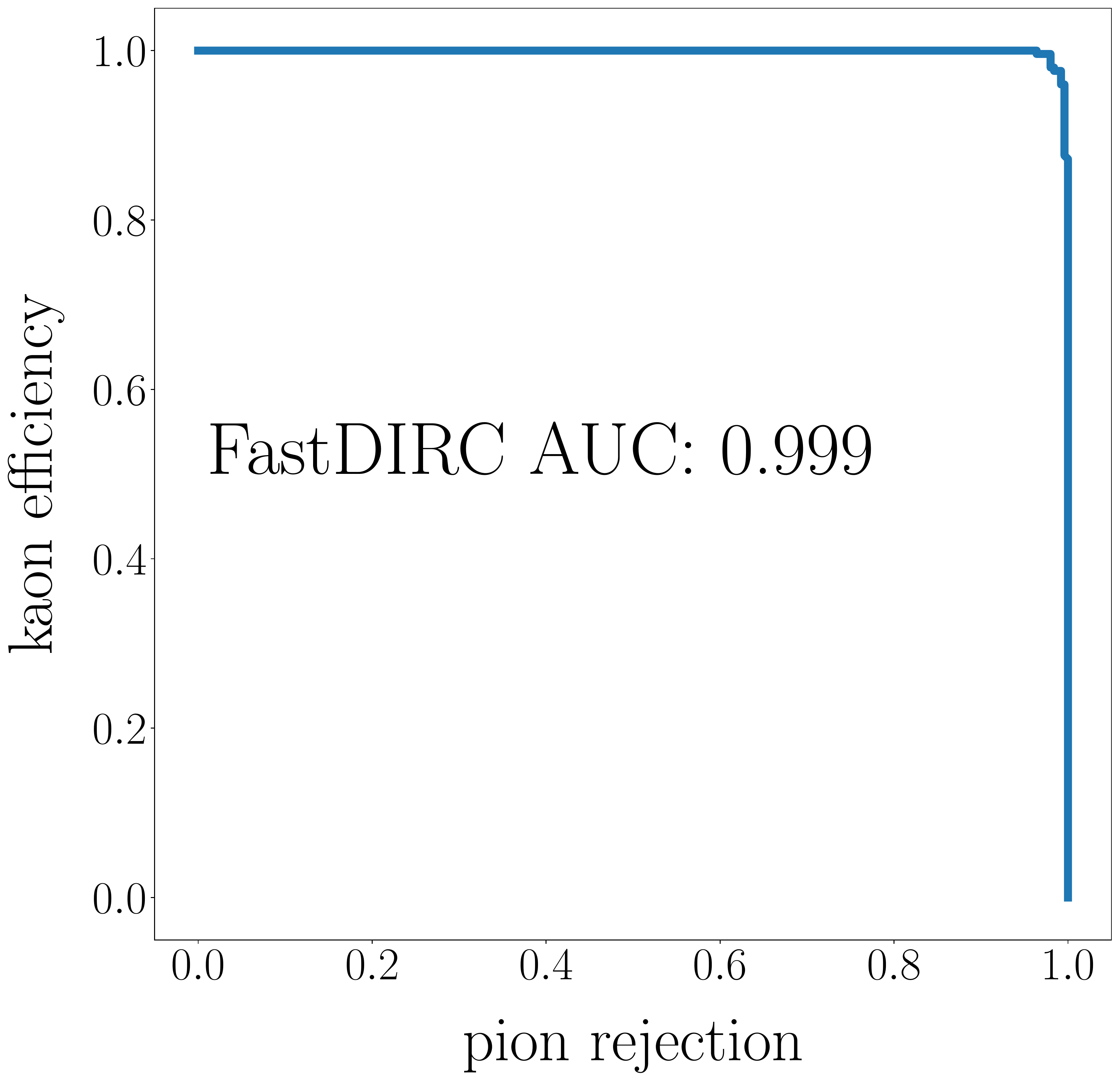}   
    \includegraphics[width=.35\textwidth,height=.35\textwidth]{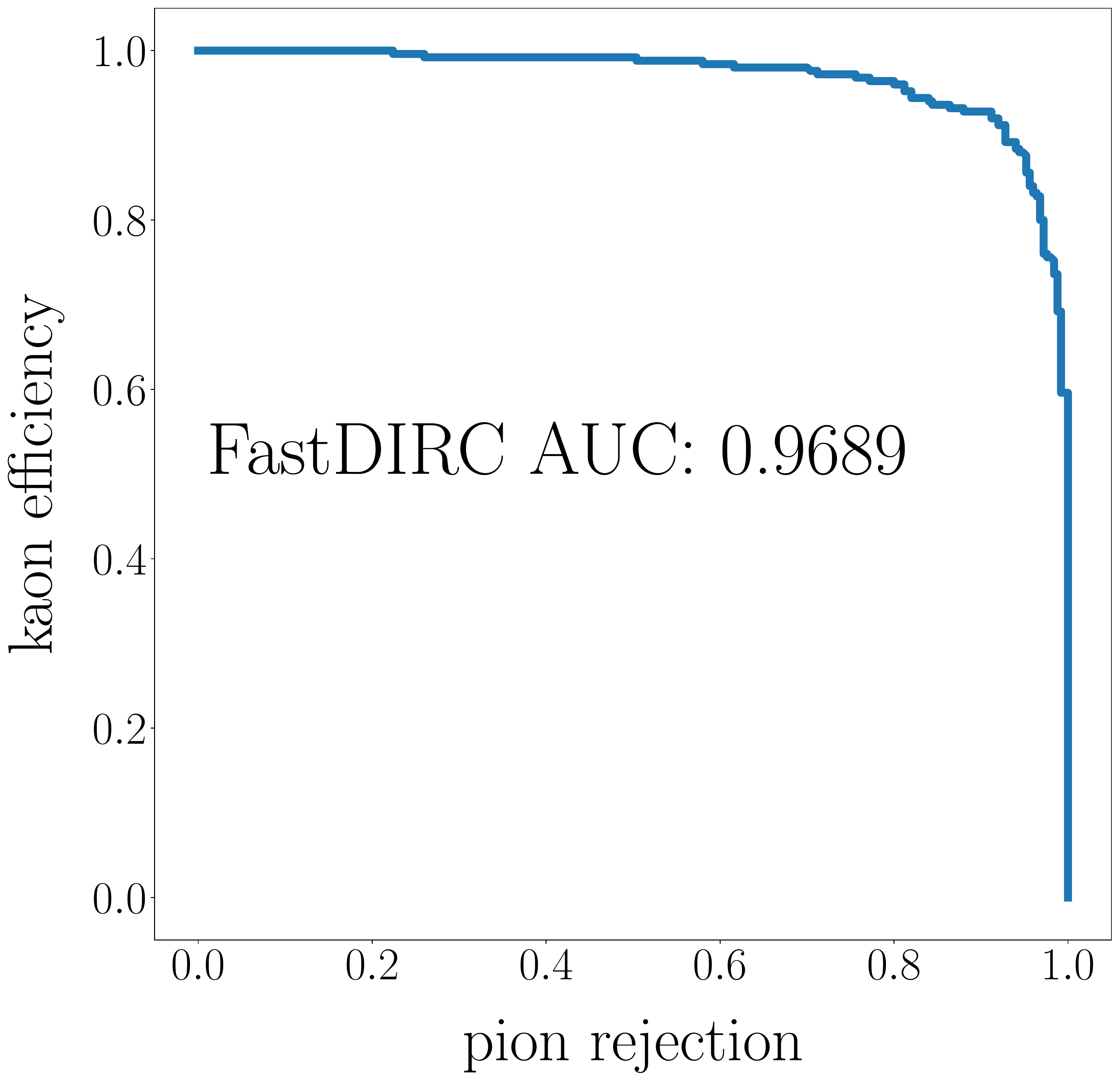}  \\
    \caption{The results are obtained with $p=$ 4 GeV/c (left column) and $p=$ 5 GeV/c (right column), at $\theta=$ 2.0 [deg], $\phi=$ 20 [deg], X = -1 [cm], Y = 55 [cm]. (top row) The overlaid distributions of $\Delta\log{\mathcal{L}}$ \ (see text for definition) for DeepRICH and FastDIRC, from which the corresponding ROC curves are calculated for comparison: \\(middle) DeepRICH; (bottom) FastDIRC.}
    \label{fig:all_gevdll_results}
\end{figure*}

At this point we can consider the $\Delta\log{\mathcal{L}}$, the difference between the two log-likelihoods (under the null hypotheses of $\pi$ and $K$, respectively). 
Histograms of $\Delta\log{\mathcal{L}}$ \ are obtained for both FastDIRC and DeepRICH and shown in Fig. \ref{fig:all_gevdll_results} at 4 GeV/c (left column) and 5 GeV/c (right column), respectively. Two different colors are used in the legend to highlight the ground truth of each particle (which is either a real $\pi$ or $K$).

In the same figures, to quantify the performance of the two algorithms, a Receiver Operating Characteristic (ROC) curve is obtained by changing the threshold on the $\Delta\log{\mathcal{L}}$ \ to cut on. 

The ROC curves have been produced generating 350 particles observed for each kinematics and the Area Under Curve (AUC) is used as a metric to compare the performance of the two algorithms.

\begin{figure*}
    \centering
    \includegraphics[width=.35\textwidth]{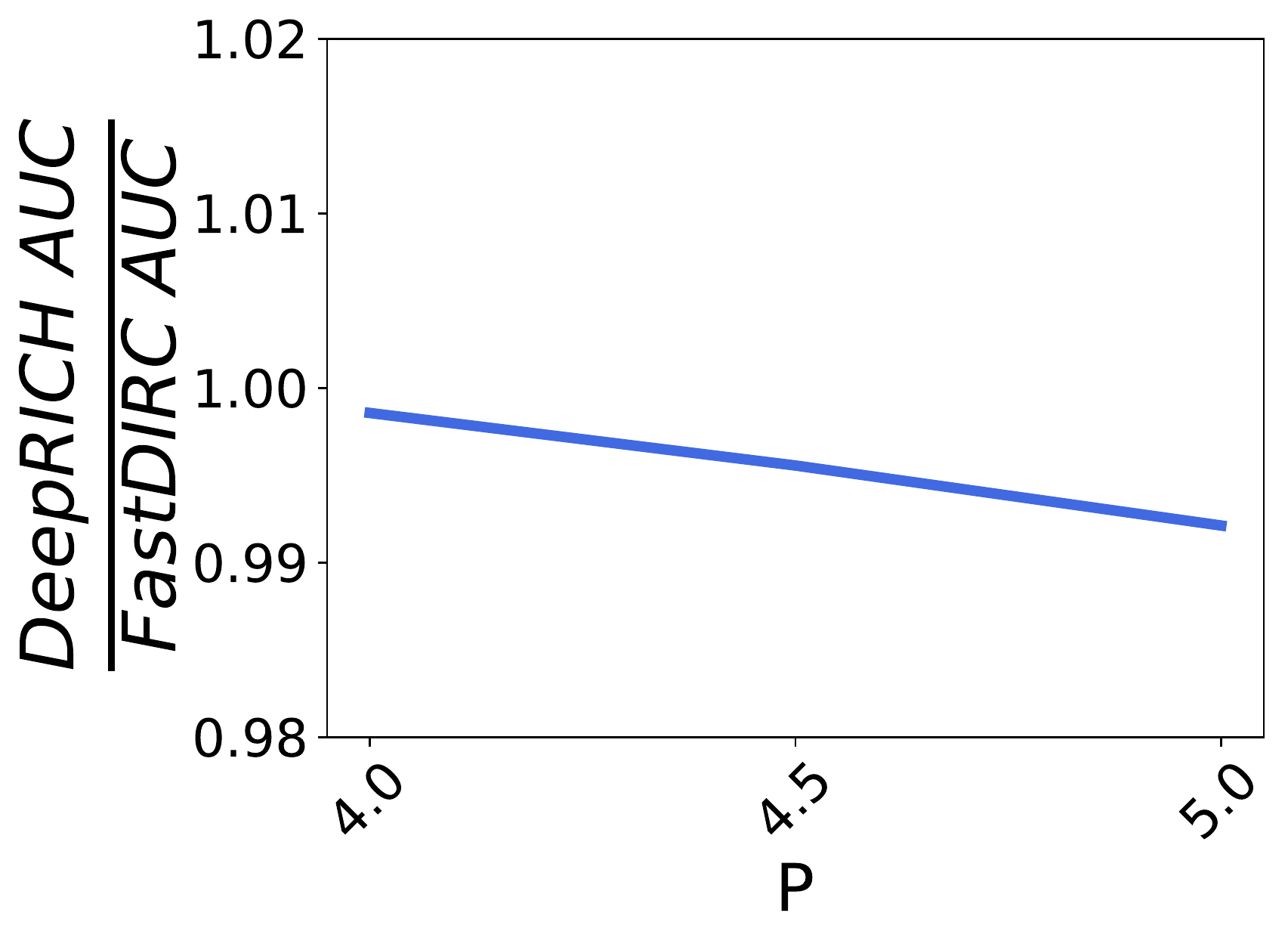} \\
    \includegraphics[width=.35\textwidth]{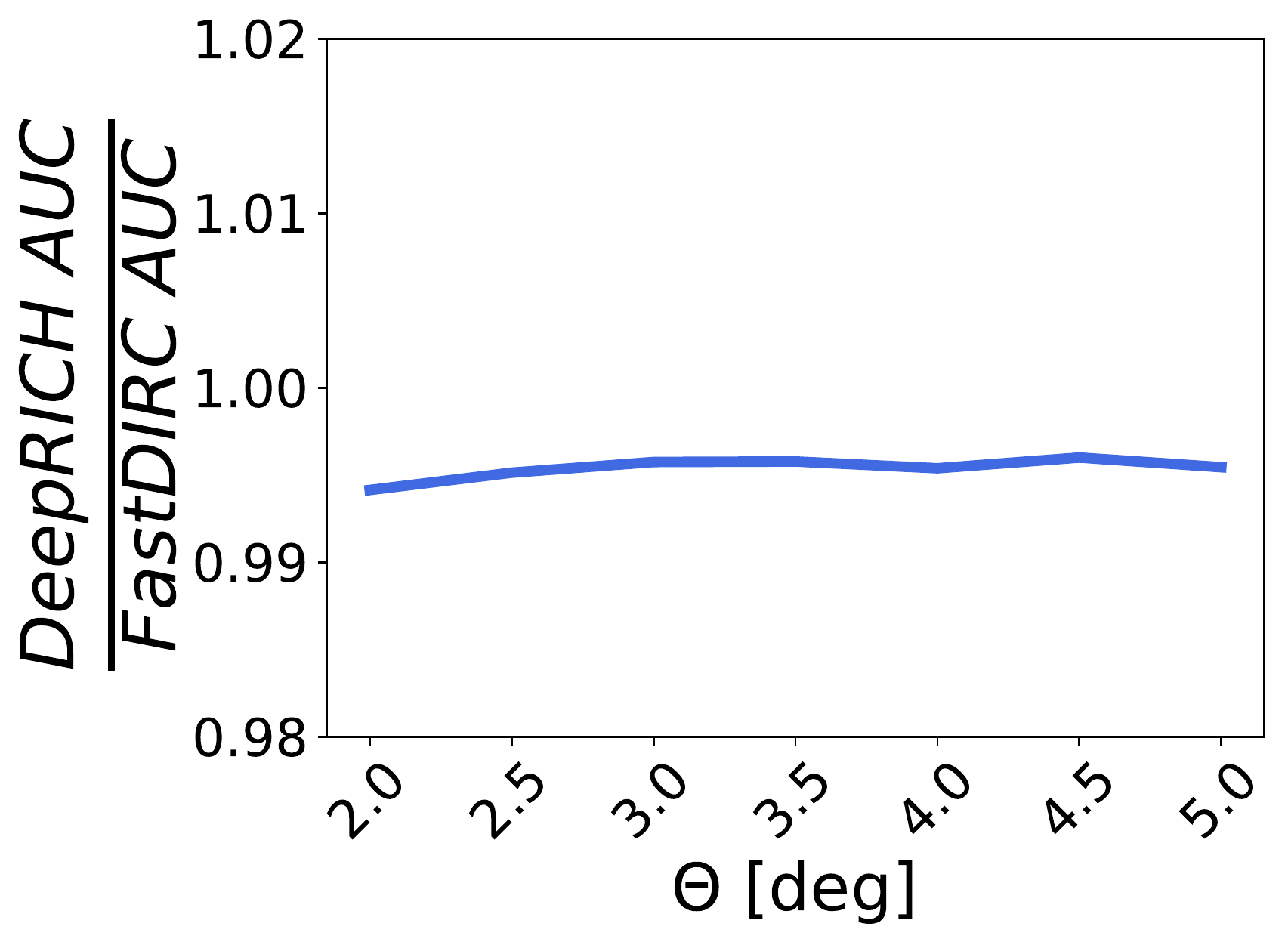}
    \includegraphics[width=.35\textwidth]{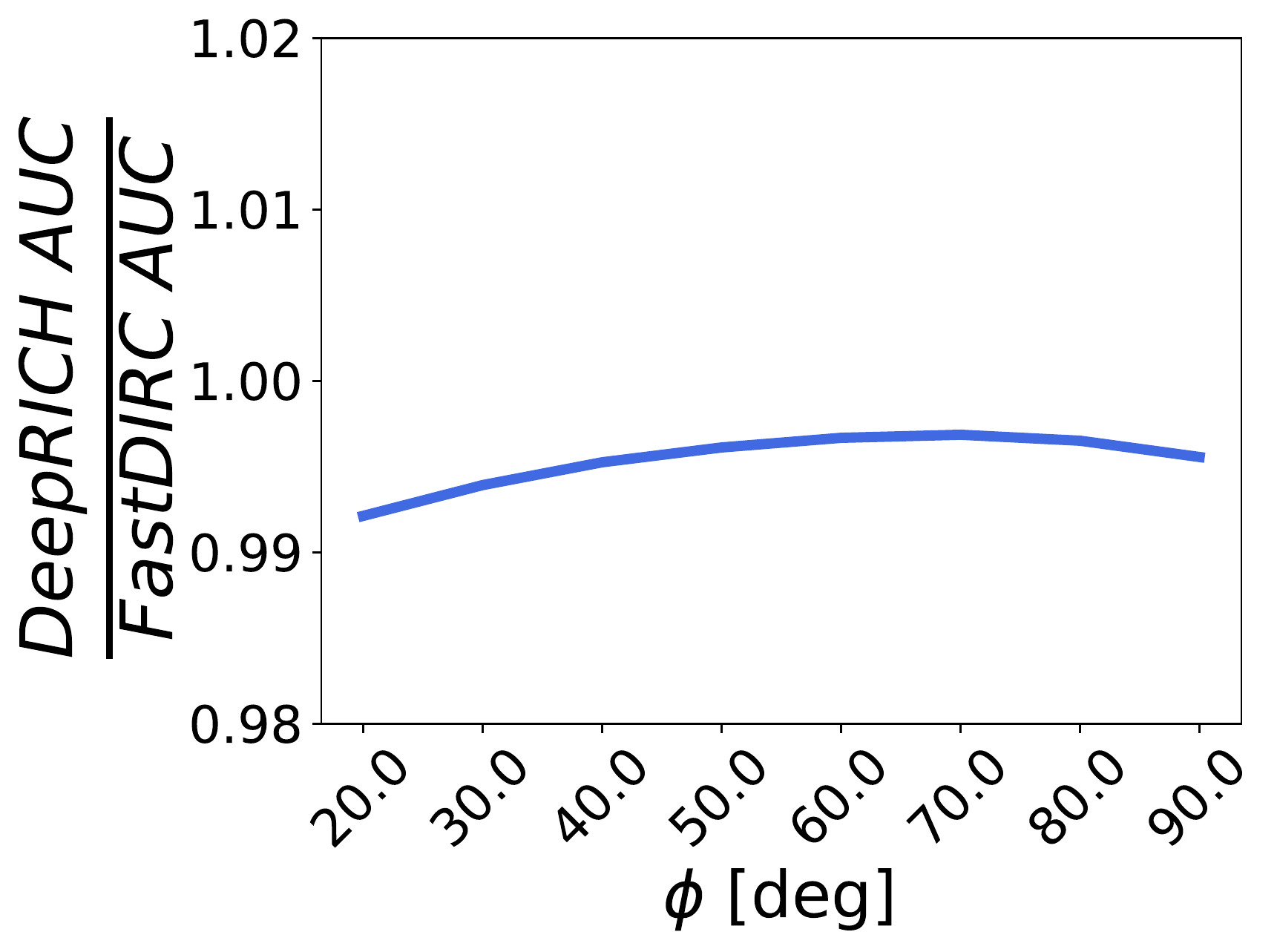}
    \includegraphics[width=.35\textwidth]{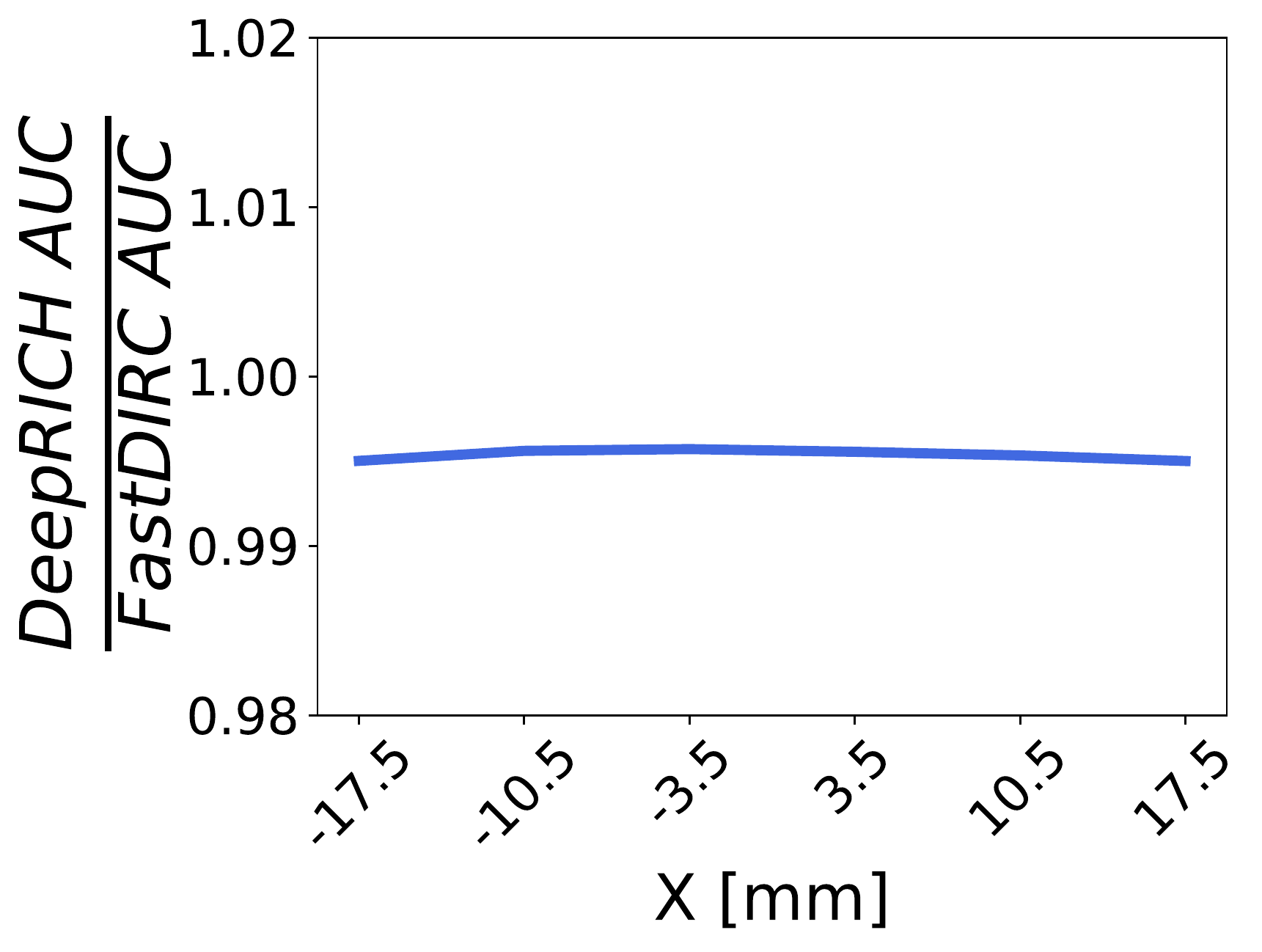}
    \includegraphics[width=.35\textwidth]{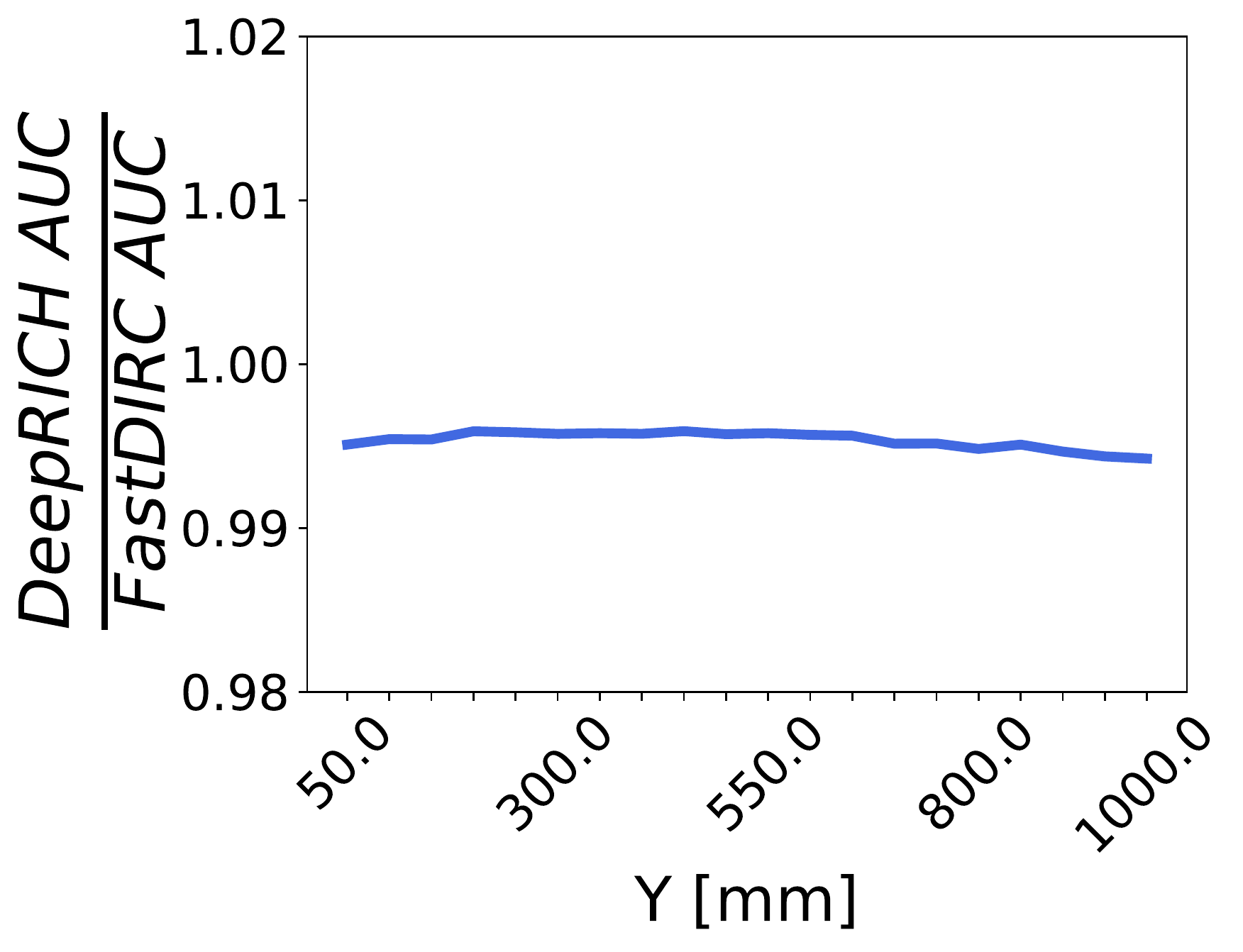}
    \includegraphics[width=.67\textwidth]{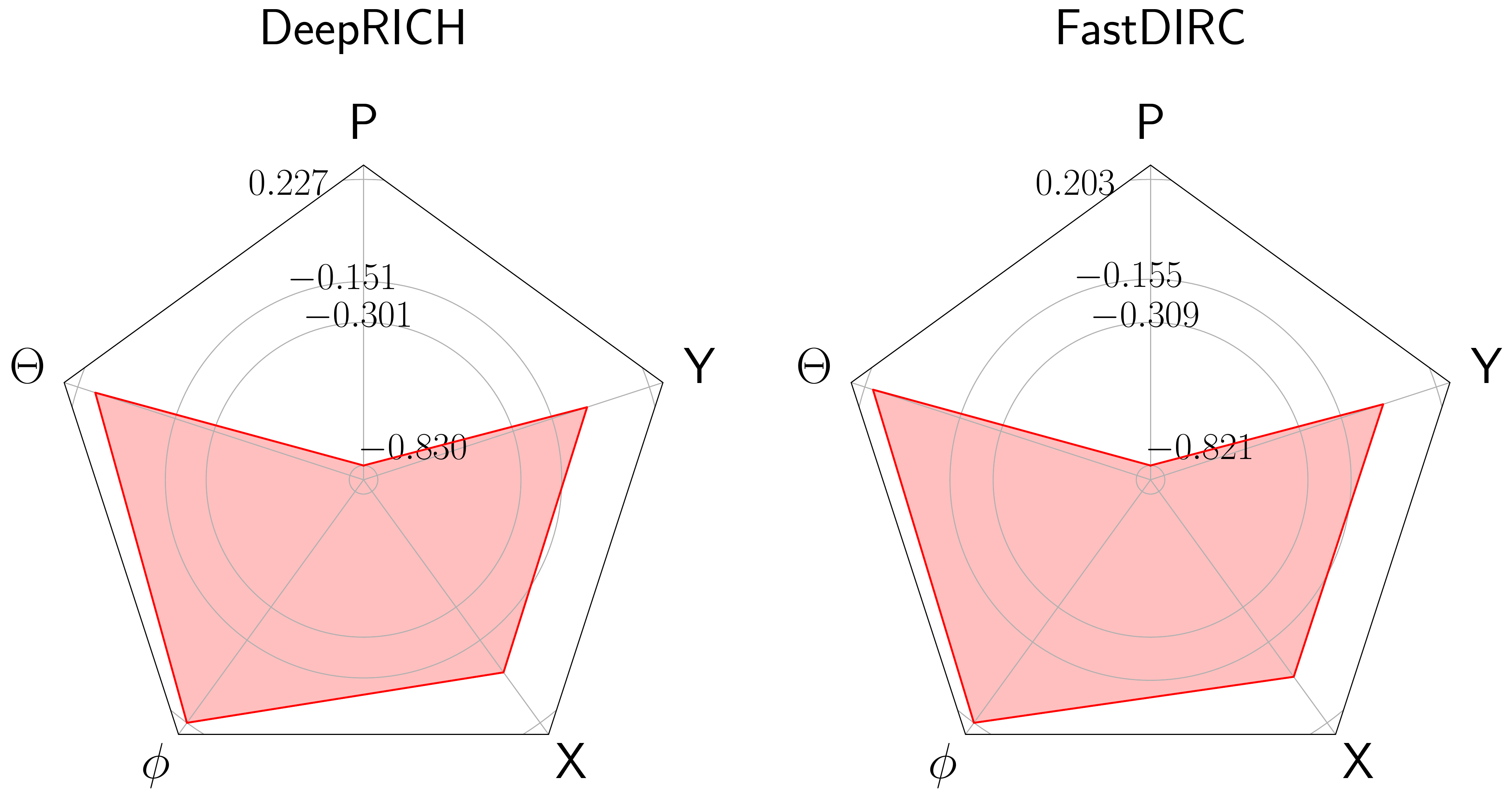}\\    
    \caption{(top 3 rows) The ratio between DeepRICH and FastDIRC AUCs. Each AUC is calculated to show the partial dependence on one kinematic parameter by marginalizing on all other parameters.
    (bottom row) Radar plots representing the correlation between the AUC and each kinematics parameter for DeepRICH and FastDIRC, showing that the two reconstruction algorithms perform similarly as a function of the kinematic parameters. Notice at 4 GeV/c that the two reconstruction methods perform almost identically.}
    \label{fig:auc_comparison}
\end{figure*}

A detailed comparison between FastDIRC and DeepRICH reconstructions is reported in
Fig.~\ref{fig:auc_comparison} (top), where the DeepRICH AUC divided by the corresponding AUC of FastDIRC is drawn as a function of a single kinematic variable, after integrating the performance over all the other kinematic parameters to show the partial dependence on that particular variable.  %

The plots show that the two algorithms are very close in reconstruction performance, namely AUC(deepRICH)~$\gtrsim$~0.99$~\cdot$~AUC(FastDIRC) in a large region of the kinematic parameters where the reconstruction efficiency of DeepRICH is approximately uniform, while a slight dependence is observed as a function of the momentum. 
Fig. \ref{fig:auc_comparison} (bottom) summarizes these results in form of radar plots: each axis correspond to a kinematic parameter, and the distance from the center on each direction corresponds to the correlation of the AUC with that specific parameter. As expected, the largest dependence of the AUC is on the momentum parameter, the $\pi$/K distinguishing power becoming lower at larger values of the momentum.

\subsection{Test on Unknown Kinematics}

One major concern about this method regards the predictability for kinematics not explicitly injected in the training phase. 
In this section we show results that prove the stability of the network reconstruction for every kinematic point belonging to the hypercube $\Delta p \times \Delta \theta \times \Delta \phi \times \Delta x \times \Delta y$, which was approximated in Sec. \ref{subsec:data} by a discrete grid of training datasets.
This approximation tacitly assumes no discontinuities in the hit pattern by varying the parameters within the hypercube. 

\begin{figure*}[ht!]
    \centering
    \includegraphics[width=0.9\textwidth]{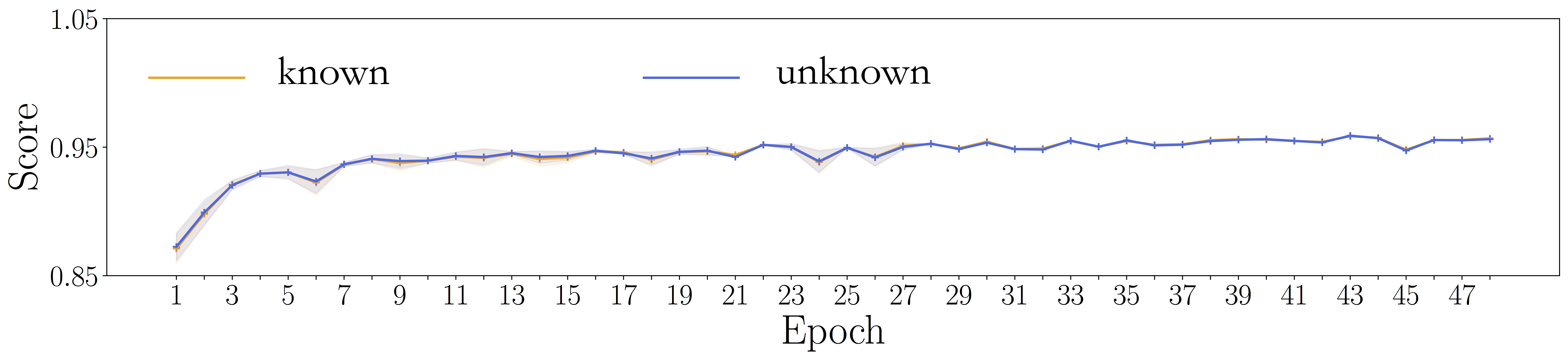}    
    \caption{The image shows the learning curve corresponding to known and unknown kinematics combining the datasets with momentum $\in$ [4,5] GeV/c. Each point is obtained as an average over 3 experiments---notice in some experiment the early stopping activated earlier. These results prove the ability of DeepRICH to reconstruct unknown kinematics.
    }
    \label{fig:uknown_kinematics}
\end{figure*}

In Fig. \ref{fig:uknown_kinematics} we show the quality of the DeepRICH reconstruction for unknown kinematics in terms of the test score. 
We performed different tests and we did not notice any sensible changes in the test score and in the AUC, which are two figures of merit we have used to prove the quality of the reconstruction. 

\subsection{DeepRICH Performance}

In this section we summarize the performance of the network both in terms of reconstruction efficiency and computing time. 
The quality of the reconstruction is high as shown in Table~\ref{tab:AUC_summary}: as already mentioned, the AUC values are close to those of FastDIRC, given a certain sub-region of the kinematic space for the training process. 
Notice these results can further improve considering the major points addressed in Sec.~\ref{subsec:data}-\ref{sec:optimization} for the training phase. 

\begin{figure*}[ht!]
    \centering
    \includegraphics[width=.7\textwidth]{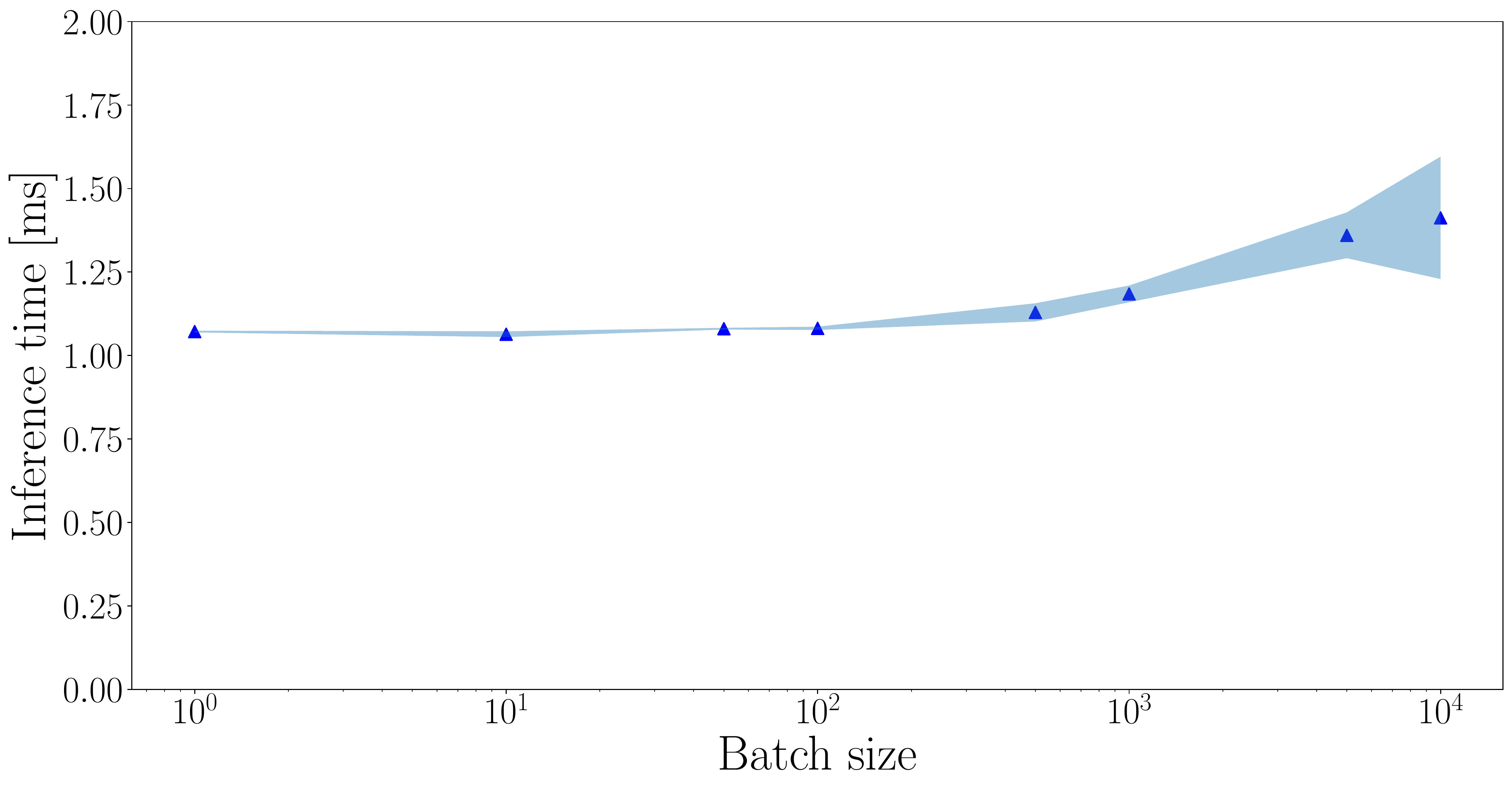}
    \caption{After training, the inference time is almost constant as a function of the batch size, meaning that the effective inference time---\textit{i.e.}, the reconstruction time per particle---can be as small as few $\mu s$. Notice that the corresponding memory size in the inference phase is approximately and equal to the value reported in Table~\ref{tab:performance}.
    }
    \label{fig:inference_vs_batch}
\end{figure*}

\begin{table}[ht]
\caption{\label{tab:AUC_summary}
The table shows in (\%) the area under curve, the signal efficiency to detect pions $\epsilon_{S}$ and the background rejection of kaons $\epsilon_{B}$ corresponding to the point of the ROC that maximizes the product $\epsilon_{S}$*$\epsilon_{B}$. The corresponding momenta at which these values have been calculated are also reported. 
This table is obtained by integrating over all the other kinematic parameters (i.e. a total of $\sim$6k points with different $\theta, \phi, X, Y$ for each momentum).
} 
\centering
\scalebox{1.0}{
\begin{tabular}{c|c|c|c|c|c|c|}
\cline{2-7}
 & \multicolumn{3}{c|}{DeepRICH} & \multicolumn{3}{c|}{FastDIRC} \\ \hline \hline
\multicolumn{1}{|l|}{Kinematics} & AUC & $\epsilon_{S}$ & $\epsilon_{B}$ & AUC & $\epsilon_{S}$  & $\epsilon_{B}$\\ \hline \hline
\multicolumn{1}{|l|}{4 GeV/c}  & 99.74  & 98.18  & 98.16 & 99.88  & 98.98  & 98.85  \\ \hline
\multicolumn{1}{|l|}{4.5 GeV/c}  & 98.78  & 95.21 & 95.21 & 99.22  & 96.33   & 96.32 \\ \hline
\multicolumn{1}{|l|}{5 GeV/c}& 96.64 & 91.13  & 91.23  & 97.41  & 92.40  & 92.47  \\ \hline
\end{tabular}
    }
\end{table}

Table~\ref{tab:performance} reports on the memory and computing time performance: the inference time is the actual time DeepRICH needs to do PID after the training phase and is on average $\mathcal{O}($ms$)$ per batch of particles using a GPU Titan V. 
Fig. \ref{fig:inference_vs_batch} shows the inference time as a function of the batch size: the inference time is approximately constant up to 10$^{4}$ particles, which is the maximum batch size that could be handled in our configuration. 

\begin{table}[ht]
\caption{\label{tab:performance}
Table of performance of the DeepRICH architecture, reporting the average inference time, the inference memory and the training memory, \textit{i.e. } the GPU memory required by the network during the inference and training phases with a fixed batch size. The workstation uses a GPU Titan V with the CUDA10.0\_0 build. The network has been implemented using PyTorch 1.2 \cite{paszke2019pytorch}.
} 
\centering
\scalebox{1.}{
\begin{tabular}{ | c | c |}
\hline
\textbf{specs} & \textbf{value} \\
\hline
inference time per batch & $\mathcal{O}$(1) ms \\
inference network memory & $\mathcal{O}$(1) GB \\
training network memory  & $\mathcal{O}$(4) GB \\
network memory on local storage  & $\sim$ 6 MB \\
network trainable parameters & 458592 \\
\hline
\end{tabular}
    }
\end{table}

For completeness we also report a comparison with the reconstruction time of other methods: for look-up-table-based algorithms, not fully optimized estimates provide order few ms per track on a single standard CPU \cite{dzhygadlo2019private}; for FastDIRC it is about 300 ms per track on a Macbook Air 2.2 GHz i7 and is dominated by the generation of the PDF, though it's worth reminding that can be massively parallelized; the GAN method \cite{derkach2019cherenkov} is the closest to our order of magnitude (but it regards the generation of $\Delta\log{\mathcal{L}}$ \ values) and the authors claim 1M particles generated per second.   
Another potential advantage of DeepRICH is the limited network size evaluated throughout all the training phase, which never exceeded 4 GB for different network configurations.
It's worth reminding that the network size depends mainly on the weights of the network and the gradients, rather than on the subspace of the kinematic parameters used in the training phase. 
\\This is a feature to keep in mind when comparing to the overall size of a look-up table obtained for example with the geometrical reconstruction method.

\section{Summary and Conclusions}\label{sec:summary}

The DeepRICH architecure developed in this paper shows very promising results. 
As a case study we consider the DIRC detector. 
Notice that DeepRICH is agnostic to the shape of the photon patterns, and in principle it can be trained to do PID for other imaging Cherenkov detectors.

The training set is generated with FastDIRC dividing the phase-space in a fine grid of points. We have made different tests changing the number of kinematic points in $p, \theta, \phi, X, Y$ and for one specific bar of the DIRC (we refer the reader to Sec. \ref{subsec:data} for more details on the preparation of data). \\We prove the high quality and stability of the reconstruction within the kinematic subspace. We then increased the space and kept the same dimensions of the neural network architecture, and this does not seem to affect the quality of the reconstruction. %
Notice that the generation of the hypercube and the resulting density can be further optimized in the future. 
Increasing the kinematic space and consequently the size of the dataset obviously results in larger training time and ideally this is limited only by computing resources and available time to train the network. 
It is worth reminding that the size of the network is related to the weights and the dimensions of the architecture. 
The measured inference time is approximately equal to 1 ms per batch and we find it is roughly constant up to 10$^4$ particles. 
Notice that further parallelization of the network can be explored during both the training and inference phases. 

Our conclusion is that DeepRICH, within the conditions described throughout the text, can reach the reconstruction efficiency of established algorithms and potentially outperform them in the reconstruction time.
\\The $\mathcal{O}$(ms) time performance per batch of particles makes this algorithm suitable for near real-time applications (\textit{e.g.} calibration). 
The high quality of reconstruction and the fast computing time are two compelling features of the DeepRICH algorithm, this coming at the cost of relatively long training time, as expected. 
If the latter aspect cannot be further optimized in the future, one can always use DeepRICH to characterize critical sub-regions of the phase-space, \textit{e.g.}, it can be applied to each bar separately.

DeepRICH has been designed to be easily generalized to classify other categories of particles, and the extension of the network is left for future development. 
An important feature is related to the nature of the VAE, which suggests a tempting scenario of generalizing DeepRICH to fast generation of events once the behavior in the latent space is learnt. 
Finally another suggestive application could be training DeepRICH using pure samples of identified particles from real data, this allowing to deeply learn the response of the Cherenkov detector.

\section*{Acknowledgments}

We thank E. Cisbani and M. Williams for useful comments.
This material is based upon work supported by the U.S.
Department of Energy, Office of Science, Office of Nuclear Physics under contract DE-FG02-94ER40818.
\\CF was also supported by the research grant prize awarded by the Jefferson Lab Associates. 
We gratefully acknowledge the support of NVIDIA Corporation with the donation of the Titan V GPU used for this research.


\bibliography{sample}

\end{document}